\documentclass[pdftex]{aa}
\usepackage{graphicx}
\usepackage{epstopdf}
\usepackage{natbib,amssymb,amsmath,txfonts,astro,url,mathtools}
\usepackage{nccmath}
\usepackage[version=3]{mhchem}
\usepackage[normalem]{ulem}
\usepackage{longtable,booktabs}
 \usepackage{lscape}

\usepackage{color}
\DeclareGraphicsExtensions{.pdf,.png,.eps,.gif,.ps}
\def \dix#1{\ensuremath{{\,\rm 10^{#1}}}}
\def \tdix#1{\ensuremath{{\,\times 10^{#1}}}}

\def\nH2{\hbox{$n_\mathrm{H_2}$}}
\def\kms{\hbox{km\,s$^{-1}$}}
\def\PKS1830{\hbox{PKS\,1830$-$211}}

\begin{document}
\setlength{\mathindent}{0pt}

\titlerunning {New insight on the interstellar ortho-to-para ratio of
  \ce{H2Cl+}} \title{The ortho-to-para ratio of \ce{H2Cl+}:
  Quasi-classical trajectory calculations and new
  simulations in light of new observations}
\authorrunning {R.~Le Gal et al.}

\author{R. Le Gal \inst{1}
\and C. Xie\inst{2} 
\and E.~Herbst \inst{1}
\and  D.~Talbi \inst{3}
\and H.~Guo\inst{2}
\and S.~Muller \inst{4}
}

\institute{Departments of Chemistry and Astronomy, University of
  Virginia, McCormick Road, Charlottesville, VA 22904, USA,
  \email{\url{romane.le_gal@cfa.harvard.edu}\thanks{new affiliation:
      Harvard-Smithsonian Center for Astrophysics, 60 Garden St.,
      Cambridge, MA 02138, USA}} \and Department of Chemistry and
  Chemical Biology, University of New Mexico, Albuquerque, New Mexico,
  87131, USA \and Laboratoire Univers et Particules de Montpellier,
  Universit\'e de Montpellier, CNRS, Place Eug\`ene Bataillon, 34095
  Montpellier, France \and Department of Space, Earth and Environment,
  Chalmers University of Technology, Onsala Space Observatory,
  SE-43992 Onsala, Sweden}

\date{Received 13 July 2017; Accepted 22 August 2017}

\abstract{Multi-hydrogenated species with proper symmetry properties
  can present different spin configurations, and thus exist under
  different spin symmetry forms, labeled as para and ortho for
  two-hydrogen molecules. We investigated here the ortho-to-para ratio
  (OPR) of \ce{H2Cl+} in the light of new observations performed in
  the z=0.89 absorber toward the lensed quasar \object{\PKS1830} with the
  Atacama Large Millimeter/submillimeter Array (ALMA). Two independent
  lines of sight were observed, to the southwest (SW) and northeast
  (NE) images of the quasar, with OPR values found to be $3.15 \pm
  0.13$ and $3.1 \pm 0.5$ in each region, respectively, in agreement
  with a spin statistical weight of 3:1.  An OPR of 3:1 for a molecule
  containing two identical hydrogen nuclei can refer to either a
  statistical result or a high-temperature limit depending on the
  reaction mechanism leading to its formation. It is thus crucial to
  identify rigorously how OPRs are produced in order to constrain the
  information that these probes can provide.  To understand the
  production of the \ce{H2Cl+} OPR, we undertook a careful theoretical
  study of the reaction mechanisms involved with the aid of
  quasi-classical trajectory calculations on a new global potential
  energy surface fit to a large number of high-level ab initio
  data. Our study shows that the major formation reaction for
  \ce{H2Cl+} produces this ion via a hydrogen abstraction rather than
  a scrambling mechanism. Such a mechanism leads to a 3:1 OPR, which
  is not changed by destruction and possible thermalization reactions
  for \ce{H2Cl+} and is thus likely to be the cause of observed 3:1
  OPR ratios, contrary to the normal assumption of scrambling.}

\keywords{astrochemistry -- ISM: molecules -- quasars: absorption
  lines -- quasars: individual: \PKS1830\ -- galaxies: ISM -- radio
  lines: galaxies}
\maketitle

\section{Introduction}
\label{sec:Intro}

Di-hydrogenated species, with identical hydrogen nuclei, can be found
in two different nuclear spin states: ortho, with a total hydrogen
nuclear spin $I=1$, and para, with a total hydrogen nuclear spin
$I=0$, with a $2I+1$ statistical weight for each state.  Similar
divisions into two or more spin states occur for molecules with three
or more identical hydrogen nuclei.  During the last decade,
nuclear-spin astrochemistry has gained interest owing to the numerous
ortho-to-para ratio (OPR) measurements produced especially by new
telescopes with unprecedented sensitivities and high spectral
resolution, giving access to previously elusive molecular lines.  Such
recent OPR determinations include those for \ce{H3+}
\citep{crabtree2011a}, \ce{H2O} \citep{lis2013,flagey2013},
\ce{NH3}\citep{persson2012}, \ce{NH2} \citep{persson2016}, \ce{H2S}
\citep{crockett2014}, \ce{H2O+} \citep{schilke2010,gerin2013},
\ce{CH2CN} \citep{vastel2015} and \ce{H2Cl+}
\citep{lis2010,gerin2013}. In thermal equilibrium, OPRs for species
with two identical protons range from their statistical weight limit
of 3:1 at high temperatures to infinity or zero as the temperature
decreases, depending on the symmetry of the rotational and electronic
wave functions. For example, the \ce{NH2} thermal OPR, as a function
of decreasing temperature, behaves in an opposite sense from that of
\ce{H2}, due to the $X^2B_1$ symmetry of the ground electronic state
of \ce{NH2}. OPRs are functions of the temperature only in thermal
equilibrium. As an example, in an isolated state, spontaneous
radiative ortho-para interconversion for \ce{H2} are known to be
extremely slow, $\approx \dix{13}$~yr \citep{raich1964,pachucki2008},
that is, they are much greater than the age of the Universe \citep{lique2014},
therefore OPRs were commonly believed to reflect a ``formation
temperature'' \citep{mumma1987}.

Although some of the observed OPRs were found to be consistent with
thermal values, others, such as those of water
\citep{lis2013,flagey2013}, \ce{H3+} \citep{crabtree2011a}, \ce{NH3}
\citep{persson2012}, and \ce{NH2} \citep{persson2016} were not. As an
example, for the non-thermal observational values of the OPR of the
radical \ce{NH2} measured toward four high-mass star-forming regions
and found to lie, depending on the position observed along the
lines-of-sight, either below the high temperature limit of three
($2.2-2.9$) or above this limit ($\sim3.5$, $\gtrsim 4.2$, and
$\gtrsim 5.0$) \citep{persson2016}. A careful theoretical study was
necessary to explain all the results. Indeed, although the use of
nuclear-spin selection rules \citep{oka2004}, leads to the
reproduction of most of the observed OPR values below three at
reasonable temperatures \citep{legal2014a,legal2014b}, it was
necessary to identify a mechanism able to at least partially
thermalize the OPR at the very low temperatures where the thermal OPR
exceeds three and goes to infinity as the temperature goes to
0~K. This need led \cite{persson2016} to consider a process previously
omitted in models: the poorly studied \ce{H + NH2} H-atom exchange
reaction to interconvert NH$_{2}$ between its ortho and para forms,
thus increasing the OPR with decreasing temperature. This suggestion
was confirmed by \cite{legal2016} who performed quasi-classical
trajectory (QCT) calculations to show that the H-exchange reaction
\ce{NH2 + H} can indeed proceed without a barrier, and therefore be
efficient in the temperature range of interest.

An understanding of these values requires a comprehensive analysis of
the processes governing the interstellar nuclear-spin chemistry
including the formation and destruction reactions and possible
conversions of the different spin symmetries both in the gas and solid
phases. Indeed, even an understanding of what appear to be thermal
values under interstellar conditions can require a detailed analysis.
Once well understood, OPRs can afford new powerful astrophysical
diagnostics on the chemical and physical conditions of their
environments, and in particular can trace their history, provided that
the memory of a chemical process can be propagated and preserved in
the molecular level population distributions
\citep{oka2004,faure2013,legal2014a}.

\section{The case of the Chloronium ion}
\label{sec:chem}

In this paper we consider the reactions and mechanisms responsible for
the observed OPRs of \ce{H2Cl+}. This ion has been detected in a
variety of galactic sources, including diffuse and translucent
interstellar clouds, the Orion Molecular Cloud, the Orion Bar
photodissociation regions \citep{lis2010,neufeld2012,neufeld2015} and
the $z=0.89$ absorber toward the lensed quasar \PKS 1830\
\citep{muller2014b}. The measured OPR values range between 2.5 and 3.2
(although with somehow large uncertainties), roughly consistent with a
3:1 OPR.

To properly interpret these OPR values, we needed to constrain the main
reactions leading to the formation and destruction of \ce{H2Cl+} and
possible interconversion of ortho-\ce{H2Cl+} and para-\ce{H2Cl+}, such
as by the H-exchange reaction \citep{neufeld2015}:
\begin{ceqn}
\begin{align}
 \ce{ortho-H2Cl+  + H <-> para-H2Cl+ + H}.
 \label{reaction1}
\end{align}
\end{ceqn}

Because \ce{Cl} can be photoionized to produce \ce{Cl+} in diffuse gas
since its ionization potential (12.97~eV~=~299.1~kcal/mol) is just
below that of atomic hydrogen:

\begin{ceqn}
\begin{align}
\ce{Cl +  Photon -> Cl+ + e-}, 
 \label{reaction2}
\end{align}
\end{ceqn}
\ce{H2Cl+} can be produced as the result of the successive
hydrogenations \citep{neufeld2009}
\begin{ceqn}
\begin{align}
\ce{Cl+ +  H2 -> HCl+ + H},
\label{reaction3}
\end{align}
\end{ceqn}
and
\begin{ceqn}
\begin{align}
\ce{HCl+ +  H2 -> H2Cl+ + H}. 
\label{reaction4}
\end{align}
\end{ceqn}

To dominate other synthetic pathways, this reactional scheme needs the
photoionization of Cl to be efficient, as will occur in diffuse and
translucent gas but not in clouds of higher column density where UV
radiation is highly attenuated. In such media, proton transfer
reaction from \ce{H3+} and other protonating ions to Cl and HCl are
likely to be competitive; for example,

\begin{ceqn} 
\begin{align}
\ce{Cl +  H3+ -> HCl+ + H2},
\label{reaction5}
\end{align}
\end{ceqn}
which competes with reactions~(\ref{reaction2}) and (\ref{reaction3}), and
\begin{ceqn} 
\begin{align}
\ce{HCl +  H3+ -> H2Cl+ + H2}.
\label{reaction6}
\end{align}
\end{ceqn}
which competes with reaction~(\ref{reaction4}) to form the \ce{H2Cl+}
OPR.  Therefore reactions~(\ref{reaction4}) and (\ref{reaction6}) can
both contribute to the determination of the \ce{H2Cl+} OPR. It
  should be noted that the \ce{H3+ + Cl -> H2Cl+ + H} reaction is not
considered in the present study. Until recently, with an extra energy
needed to dissociate \ce{H3+} into \ce{H2+ + H}, such a process was
assumed to be non-competitive for reactions of other atoms with
\ce{H3+} \citep[e.g.][]{talbi1991}. But recent measurements show that
for the reactions \ce{C + H3+} and \ce{O + H3+}, the production of
\ce{CH2+} and \ce{H2O+} can be competitive
\citep{vissapragada2016,deRuette2016}. More work is clearly needed on
the \ce{Cl + H3+} system.

To determine the contributions of reactions~(\ref{reaction4}) and
(\ref{reaction6}) in the \ce{H2Cl+} OPR formation, we must first find
out if one reaction dominates the other in rate or whether both are
comparable. And secondly, we need to identify whether these
reactions occur via a simple hydrogen atom abstraction process for
reaction~(\ref{reaction4}) and proton hop process for
reaction~(\ref{reaction6}) or via a full scrambling process in which
both H-abstraction/proton hop and H-exchange can occur. Taking
reaction~(\ref{reaction4}) as an example, the term ``H-exchange''
refers to a process in which both H atoms on \ce{H2} end up on
\ce{Cl+} while the original H on \ce{HCl+} leaves, while
``H-abstraction'' is a simple process in which an H of \ce{H2} hops to
\ce{HCl+}. As for reaction~(\ref{reaction6}), the term ``proton hop''
is used similarly to the term ``H-abstraction'' when a proton hops
from an hydrogenated ion to the molecule with which it reacts to form
the new hydrogenated molecular ion in each of its plausible
nuclear-spin configurations. Being able to distinguish which mechanism
dominates allows one to identify the spin state(s) of the product
\ce{H2Cl+} and their expected ratio. For example, a simple hydrogen
abstraction in reaction~(\ref{reaction4}) or a simple proton hop in
reaction~(\ref{reaction6}) leads to an OPR of 3:1, while scrambling
will lead to a different value depending on the OPR of \ce{H2} for
reaction (\ref{reaction4}) and on the OPR of \ce{H3+} for
reaction~(\ref{reaction6}). Thus, the resulting formation OPR for
\ce{H2Cl+} depends upon whether {\it (i)} an intermediate complex can
be formed and {\it (ii)} its lifetime is long enough to allow the
reaction to occur through a scrambling process
\citep{oka2004,herbst2015}. Consequently, Oka's theory \citep{oka2004}
cannot be used unambiguously without detailed knowledge of the
dynamics of a chemical reaction, which requires the study of the
potential energy surface (PES) as well as trajectory studies to
determine the dominant mechanism leading to the products. In its basic
form, Oka's theory indeed imposes a full proton scrambling for all
products involving proton spin distinctions, although it can also be
used for restriction to hydrogen abstraction. The approach also
contains the assumption that the reaction is sufficiently exothermic
so that enough product rotational states can be populated to not
constrain calculated ortho-para distributions.

These different mechanisms have been explored to interpret the OPR
values of \ce{H3+} \citep{crabtree2011b,crabtree2011c} and
subsequently studied in more detail for the specific system \ce{H2 +
  H3+} \citep{gomez-carrasco2012}. The mechanistic distinctions were
also recently considered for observations of the OPRs of \ce{H2O+}
\citep{herbst2015} and \ce{H2Cl+} \citep{neufeld2015}, providing new
insights to explore. A rigorous determination of the mechanisms in a
particular system requires a full quantum scattering treatment, but as
pointed out by \cite{gomez-carrasco2012}, full quantal computations
are rarely achievable because of their complexity and the
  computational cost. The next best approach is the semi-classical
trajectory method, in which atoms and molecules undergo classical
trajectories on a quantum mechanical PES. It is this approach that is
used in this paper. Of course, both destruction and conversion
processes must also be considered in determining OPRs, and such
processes are also contained in our chemical simulations.

In the remainder of the paper, we first consider in
Sect.~\ref{sec:obs} the astronomical observations leading to new OPR
measurements for \ce{H2Cl+} in an extragalactic absorber. Then we
discuss in Sect.~\ref{sec:modeling} our modeling approach and
present its outcomes for the scrambling versus hydrogen
abstraction/proton hop mechanisms. Section~\ref{sec:theo} describes
our semi-classical treatment of reaction~(\ref{reaction4}), which
models show to be the dominant formation process for
\ce{H2Cl+}. Finally, Sect.~\ref{sec:conclu} summarizes our results
and draws the main conclusions of the present study.

\section{Observations}
\label{sec:obs}

\subsection{Description of the observations}

The ortho and para forms of \ce{H2Cl+} were observed with the Atacama
Large Millimeter/submillimeter Array (ALMA) during Early Science cycle
2, in summer 2014. Two different tunings were observed to cover the
$1_{10}$-$1_{01}$ transition of ortho-\ce{H2Cl+} (rest frequencies
$\sim 189$~GHz, redshifted to $\sim 100$~GHz, within ALMA Band 3) and
the $1_{11}$-$0_{00}$ transitions of para-\ce{H2Cl+} (rest frequencies
$\sim 485$~GHz, redshifted to $\sim 257$~GHz, within ALMA Band 6),
respectively. The array was maintained in a configuration
providing a synthesized beam resolution of better than 0.5$''$,
sufficient to resolve the two point-like lensed images of \PKS1830,
separated by 1$''$. The bandpass calibration was done on the bright
quasar 1924$-$292, and the gain calibration was solved by regular
visits on 1832$-$2039 during the observations. The correlator was set
up with spectral windows of 1.875~GHz, providing a velocity resolution
of 1.3~\kms\ (Band 6) and 2.9~\kms\ (Band 3), after Hanning
smoothing. The data calibration was performed within
CASA\footnote{http://casa.nrao.edu/} following a standard procedure. A
further run of self-calibration using the bright continuum of
\PKS1830\ helped us to significantly improve the data quality. The
final spectra were extracted toward both lensed images of \PKS1830,
using the CASA-python task UVMULTIFIT \citep{marti-vidal2014}, by
fitting a model of two point sources to the interferometric
visibilities.

The ALMA spectra of \ce{H2Cl+} toward \PKS1830\ are shown in
Fig.~\ref{fig:spec-H2Clp}. The $^{37}$-isotopologues of \ce{H2Cl+}
were also detected in the same spectral bands, with a
$^{35}$Cl/$^{37}$Cl ratio $\sim 3$. Since they have weaker absorption,
they did not bring a significant improvement on the measurement
of the OPR, and they will be reported in another publication.

\begin{table*}[ht]
\renewcommand{\arraystretch}{1.5}
\caption{H$_2$Cl$^+$ lines observed with ALMA.}
\label{tab:spectro}
\begin{center} \begin{tabular}{ccccccc}
\hline \hline
Species & o/p form & Line & Rest frequency $^{(a)}$ & Sky frequency $^{(b)}$ & Dates of observations \\
        & &     & (GHz)      &  (GHz) &               \\
\hline
H$_2$Cl$^+$ & ortho &$1_{10}$-$1_{01}$ & 189.225 $^{(c)}$ & 100.341 & 2014, Jul. 21 and Aug. 26  \\ 
H$_2$Cl$^+$ & para & $1_{11}$-$0_{00}$ & 485.418 $^{(c)}$ & 257.404 & 2014, Jun. 06 and Jul. 29  \\
\hline
\end{tabular} 
\tablefoot{ $a)$ Rest frequencies are taken from the Cologne Database
  for Molecular Spectroscopy (CDMS,
  http://www.astro.uni-koeln.de/cdms/) and references therein.  $b)$
  Sky frequencies are given assuming a redshift $z_{abs}=0.88582$,
  heliocentric frame.
$c)$ The frequency is that of the strongest hyperfine component.
}
\end{center} 
\end{table*}

\subsection{Description of the source}

The quasar \PKS1830\ is lensed by a nearly face-on typical spiral
galaxy \citep[e.g.][]{wiklind1998,winn2002,koopmans2005}, and appears
as two bright and compact images separated by 1 arcsec, seen at mm
wavelengths, and embedded into a pseudo-Einstein ring only seen at cm
wavelengths \citep{jauncey1991}. Molecular absorption arises along the
two lines of sight toward the two lensed images, at projected
galactocentric radii of $\sim 2$~kpc for the southwest (SW) image
and $\sim 4$~kpc for the northeast (NE) image, in opposite
directions from the center of the absorbing galaxy, respectively. The
absorption system has been observed at mm and submm wavelengths
\citep{muller2011,muller2014a} leading to the detection of more than
forty different molecular species toward the SW image. The \ce{H2}
column density is $\sim 2 \times 10^{22}$~\cc\
\citep{muller2011,muller2014a} in this line of sight. In contrast,
only about a dozen species have been detected on the NE line of sight,
where the \ce{H2} column density is $\sim 1 \times 10^{21}$~\cc,
  that is to say a factor 20 lower.

Molecules tracing gas with a low fraction of molecular hydrogen
($f_{\rm H_2}$) preferentially, such as ArH$^+$, OH$^+$, H$_2$O$^+$,
and H$_2$Cl$^+$, are found to have enhanced abundances along the NE
line of sight relative to the SW one
(\citealt{muller2014b,muller2015,muller2016a}). Among other evidence,
this observation suggests that the absorbing gas in the NE line of
sight is more diffuse. The cosmic-ray ionization rate of atomic
hydrogen was also investigated using the abundance ratio of OH$^+$ and
\ce{H2O+} \citep{muller2016a}, leading to a ratio of about one order
of magnitude between the two sightlines.

For the SW absorption, the physical conditions were constrained
by \cite{henkel2008,henkel2009} using \ce{NH3} and \ce{HC3N} and by
\cite{muller2013} using multi-transition observations of a set of
various molecular species (\ce{CH3CN}, SO, \ce{c-C3H2}, \ce{HC3N},
H$^{13}$CO$^{+}$, H$^{13}$CN, HC$^{15}$N, HNCO, and \ce{SiO}). The
absorbing gas has properties similar to Galactic diffuse to translucent
clouds, with a kinetic temperature of $\sim 80$~K and an H$_2$ volume
density of the order of $10^3$~cm$^{-3}$. The temperature and density
determined by \cite{muller2013} might be only relevant to the peculiar
absorbing component traced by their observed molecules. It is also
likely that these conditions are average values over the volume of
absorbing gas, which is roughly encompassed in a pencil beam with a
base of $\leq$ 1 pc in diameter and a length of few tens to hundreds
of pc.  See also \cite{schulz2015} and \cite{muller2016b} for evidence
of inhomogeneities in the absorbing gas.

For the NE line of sight, the conditions are only poorly
constrained. There is evidence for a more diffuse gas component than
for the SW line of sight. Therefore, the two lines of sight present
clearly different physical and chemical gas properties, which makes
all the more interesting the comparison of the observed H$_2$Cl$^+$
OPR between the two.

 \subsection{Observed absorption spectra}

 \begin{figure}[!h] 
\centering
     \includegraphics[width=8.8cm]{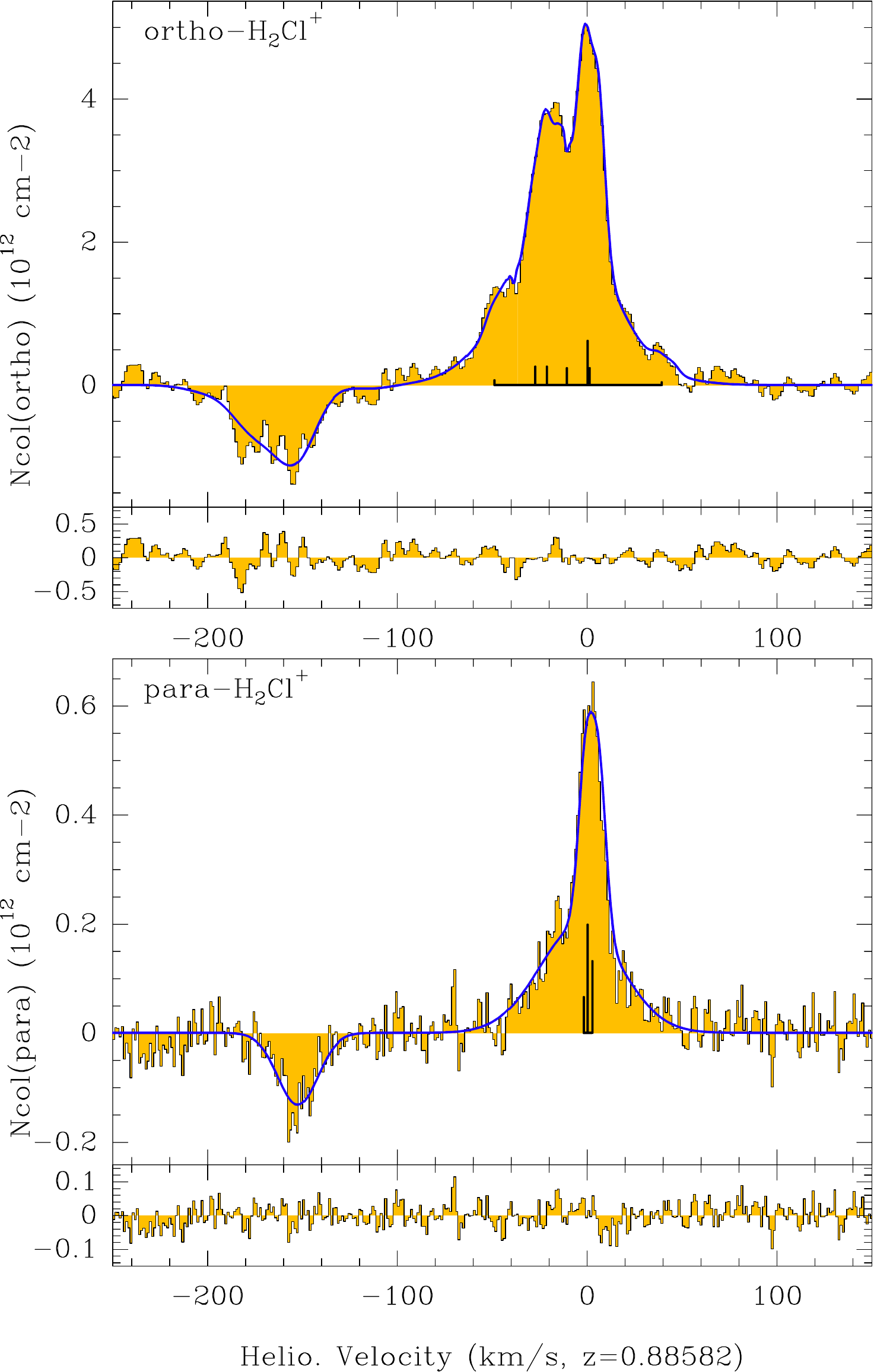}
     \caption{Spectra of the ortho (upper panel) and para (lower
       panel) forms of \ce{H2Cl+} toward \PKS1830. The y-axis is in
       unit of column density.The opacities were converted into column
       densities assuming that the excitation is locked to the
       temperature of the CMB.  The spectra of the SW line of sight
       have been divided by that of the NE. The SW absorption appears
       as positive feature near $v=0~\kms$, while the NE absorption
       appears as the negative feature near $v=-150~\kms$. The fit
       with five Gaussian velocity components, including both ortho
       and para forms, is shown in blue on top of the spectras, and
       the corresponding residuals are shown in the lower boxes. The
       velocity scale is set for the strongest hyperfine
       component. The hyperfine structure is plotted at the base of
       each spectrum.}
     \label{fig:spec-H2Clp} 
\end{figure}

The detection of the para-\ce{H2Cl+} $1_{11}$-$0_{00}$ line along both
the SW and NE lines of sight has already been reported by
\cite{muller2014b}. Assuming for each sightline that 3/4 is
contributed by the ortho \ce{H2Cl+} and 1/4 by the para \ce{H2Cl+},
they estimated total \ce{H2Cl+} column densities of $1.4 \tdix{13}
\cc$ in the SW line of sight, and $4 \tdix{12} \cc$ in the NE line of
sight.

With the new ALMA data reported here of both the ortho- and
para-\ce{H2Cl+} forms, we can now measure the OPR of \ce{H2Cl+} in the
absorber toward \PKS1830. We perform a simultaneous fit of the ortho-
and para-\ce{H2Cl+} absorption profile with five Gaussian velocity
components (four for the SW line-of-sight and one for the
NE one), with the OPR as a free parameter of the fit. The
corresponding fit is shown in Fig.~\ref{fig:spec-H2Clp}.  We
  obtained $\rm OPR_{\rm SW} = 3.15 \pm 0.13$ and $\rm OPR_{\rm NE} =
3.1 \pm 0.5$, respectively.

\section{Modeling}
\label{sec:modeling}

The interstellar chemistry of chlorine was first discussed by
\cite{jura1974} and \cite{dalgarno1974}, who identified the most
important reactions in the ion-molecule scheme and showed that
\ce{H2Cl+} is one of the species produced on the way to HCl. More
recently, the study of interstellar chloronium chemistry was
re-investigated by \cite{neufeld2009}, who developed a gas-phase
chemical network that they coupled with the PDR Meudon Code
\citep{lepetit2006} to provide, among other chlorine-bearing species,
\ce{H2Cl+} abundance predictions for diffuse and translucent
clouds. This study and the following study by \cite{neufeld2012}
mainly focused on the total amount of \ce{H2Cl+} and did not consider
the different nuclear-spin configurations of \ce{H2Cl+}. In the light
of new observations, \cite{neufeld2015} discussed the different
mechanisms that could drive the \ce{H2Cl+} OPR. They focused their
discussion on the main formation reaction, reaction (\ref{reaction4}),
and showed how the \ce{H2Cl+} OPR should depend on the OPR of \ce{H2}
if a full scrambling mechanism prevails, while a constant value of 3:1
is obtained with the hydrogen abstraction mechanism. They then showed
that the observational data were more consistent with a hydrogen
abstraction mechanism, but did not undertake a detailed study of the
different plausible mechanisms for reaction (\ref{reaction4}). Nor did
they consider the possibility that reaction (\ref{reaction6}) can also
contribute to the OPR of \ce{H2Cl+}. With our nuclear-spin chemical
network -- which is the first to our knowledge to consider the
different nuclear-spin configurations for \ce{H2Cl+} -- and our
detailed quantum chemistry and trajectory studies, we have included
both of these avenues of research.

\subsection{Model description}

We enlarged the nuclear spin chemical network used in
\cite{legal2016}, which includes the hydrogen nuclear-spin chemistry
as well as those of other species, with the addition of the chloronium
nuclear-spin chemistry that we developed for this work, based on the
theory of \citet{oka2004}. Except for the newly added \ce{H2Cl+} spin
chemistry, all reactions with distinct spin forms considered in the
chemical network proceed via scrambling. For the \ce{H2Cl+} OPR, we
included the influence of the different plausible mechanisms, hydrogen
abstraction and proton hop versus full scrambling for both reaction
(\ref{reaction4}) and reaction (\ref{reaction6}), respectively. In
addition, we assumed that the destruction reactions for \ce{H2Cl+} are
not dependent on its spin state. To run our simulations, we coupled
our network to the \textsc{Nautilus} code \citep{ruaud2016}, used in
the gas-phase mode, to run pseudo-time-dependent simulations with the
use of typical diffuse ISM gas-phase elemental abundances relative to
total hydrogen \citep{gerin2016}, as listed in
Table~\ref{tab:initial_ab}.  Each pseudo-time-dependent simulation was
run with constant physical conditions. The gas was considered shielded
from the interstellar radiation field \citep[ISRF][]{mathis1983} by
one to three magnitudes of visual extinction with a fixed cosmic ray
ionization rate of hydrogen that we set equal to the values estimated
by \cite{muller2016a}: $\zeta_\ce{H2} \approx 3 \tdix{-14} \pers$ for
the SW region and $\zeta_\ce{H2} \approx 5 \tdix{-15} \pers$ for the
NE region using the scaling relation 2.3$\zeta_\ce{H} \approx 1.5
\zeta_\ce{H2}$ from \cite{glassgold1974}. The only processes involving
dust grains are charge exchanges, dissociative recombinations, and the
formation of \ce{H2}. We adopted a single dust grain radius of
0.1~$\mu$m, satisfying a dust-to-gas mass ratio of 1\%.

\begin{table}
\renewcommand{\arraystretch}{1.5}
  \tiny
  \centering
  \caption{Elemental abundances and initial species used in this
    work.}
\begin{tabular} {lcc}
  \hline\hline
  Species    &  Elemental abundances$^{(a)}$ \\
  \hline
  \ce{He}     &     1.00(-1)  \\ 
  \ce{N}       &     6.80(-5)  \\ 
  \ce{O}       &     3.10(-4)  \\
  \ce{H}         &     0.96      \\ 
  \ce{H2}       &     0.02      \\
  \ce{C+}     &     1.40(-4) \\
  \ce{S+}     &     1.40(-5) \\
 \ce{Fe+}   &    3.00(-9) \\
 \ce{Cl+}    &    1.80(-7) \\
 \hline
\end{tabular}
\tablefoot{\tiny Numbers in parentheses are powers of ten. Elemental abundances with respect to the total proton abundance. The electron abundance is computed
  internally in the code in order to have a neutral gas.
  \tablefoottext{a}{Elemental abundances
    from \cite{gerin2016}.}}
\label{tab:initial_ab}
\end{table}

\subsection{Modeling results}
\label{subsec:modeling}

\begin{figure}[h] 
\centering
     \includegraphics[scale=0.44]{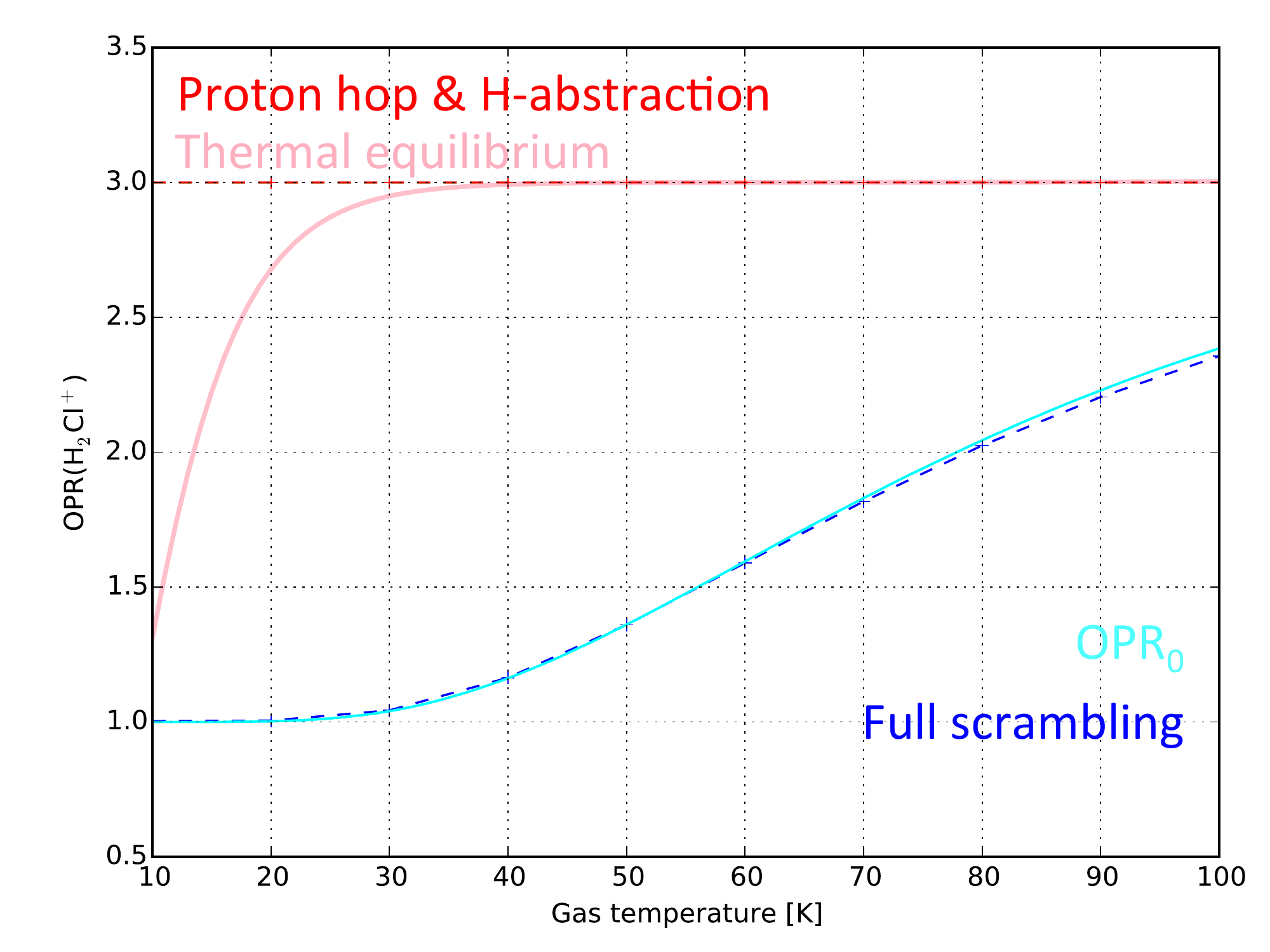}
     \includegraphics[scale=0.44]{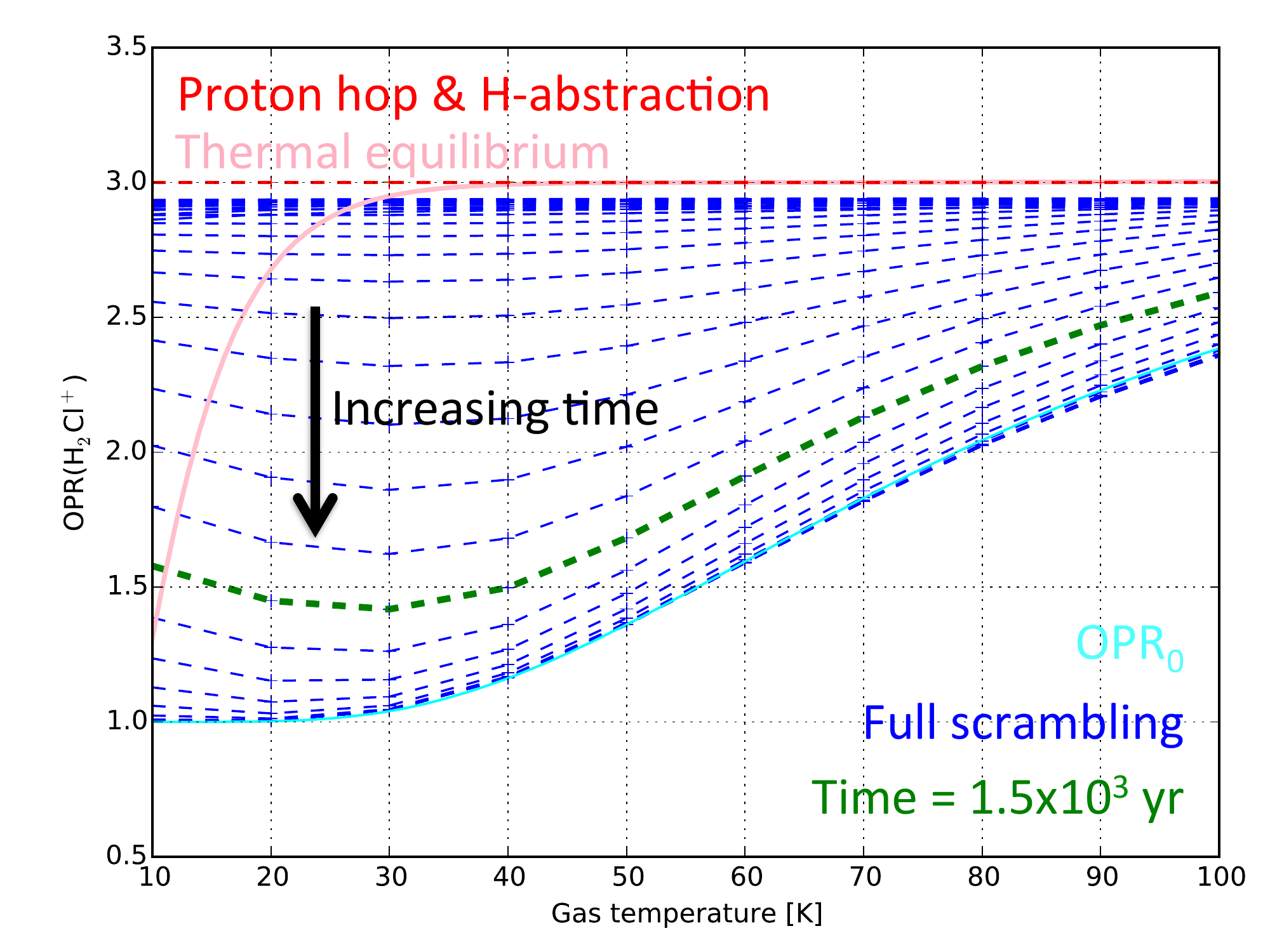}
     \caption{\ce{H2Cl+} OPR as a function of the gas temperature:
       {\it (i)} at thermal equilibirum (solid pink line), {\it (ii)}
       obtained with a full model considering a hydrogen abstraction
       and a proton hop mechanism for respectively the \ce{H2 + HCl+}
       and \ce{H3+ + HCl} formation reactions (dashed red curve), {\it
         (iii)} obtained with a model considering full scrambling for
       all the reactions involving nuclear-spin forms (dashed blue
       curves), {\it (iv)} taken from a simple analytical formula and
       designated OPR$_{0}$, this model represents scrambling via
       reaction (\ref{reaction4}) at steady state with the OPR for
       H$_{2}$ treated as thermal \citep{neufeld2015} (solid
       cyan). The time evolution of the complete scrambling model is
       depicted on the second panel where the \ce{H2Cl+} OPR is
       represented in green for a time of $1.5\tdix{3}$~yr as an
       indicative reference.}
     \label{fig:model-OPR-H2Clp} 
\end{figure}

To test the impact of the physical conditions on \ce{H2Cl+} formation,
we ran grids of models where the density, $n_\ce{H} = n(\ce{H}) +
2n(\ce{H2})$, varies from $2\tdix{2}$ to $2\tdix{4}$~\ccc\ and the gas
temperature from 10 to 100~K. The results show that the density does
not strongly affect the \ce{H2Cl+} OPR but that the temperature does
depending on the mechanism considered, as displayed in
Fig.~\ref{fig:model-OPR-H2Clp}. The figure represents the \ce{H2Cl+}
OPR as function of the temperature for four different models.  In
addition, the time dependence of the models is designated in the lower
panel. The thermal equilibrium value is plotted on the figure, as is
the result of our detailed model in which the hydrogen abstraction and
proton hop mechanisms are considered for the \ce{H2 +HCl+} and \ce{H3+
  + HCl} formation reactions, respectively. We also plot the results
of our detailed model in which full scrambling is considered for all
formation reactions involving nuclear spin forms. This model is the
only one with noticeable time dependence.  The derived curve at steady
state is almost identical to a much simpler model, designated
OPR$_{0}$ in the figure, in which an analytical formula was used to
express the OPR of \ce{H2Cl+} in terms of the thermal OPR of H$_{2}$
by including only reaction~(\ref{reaction4}), which occurs via
scrambling \citep{neufeld2015}. The excellent agreement at
steady-state between these two latter models shows that the importance
of reaction~(\ref{reaction6}) is negligible. For the full scrambling
model, a time of $1.5\tdix{3}$~yr is emphasized in the figure as an
indicative reference to show how fast the \ce{H2Cl+} OPR reaches its
steady state value. Thus we clearly see that an \ce{H2Cl+} OPR of 3:1,
even at a temperature as high as 80~K, is excluded with the
consideration of a full scrambling mechanism.

Regarding the full scrambling results it might be legitimate to think
that once the OPR of \ce{H2Cl+} is formed, subsequent thermalization
processes could occur as the para-to-ortho interconversion of
reaction~(\ref{reaction1}), as suggested in \cite{neufeld2015}. Such a
mechanism would lead to thermal equilibrium if it could occur rapidly
enough.  To address this question, we studied the energetics of this
reaction in order to investigate whether \ce{H2Cl+} can react with H
atoms to interconvert and thermalize the \ce{H2Cl+} OPR, similar to
our previous calculation on \ce{NH2 + H} \citep{legal2016} in which
NH$_{2}$ is driven toward thermal equilibrium. For the \ce{H2Cl+}
case, however, a high barrier was found for this interconversion, as
is discussed in Sect.~\ref{sec:theo}. The reaction \ce{H2 + p-H2Cl+
  <-> H2 + o-H2Cl+} is even less plausible. If the reaction occurs via
H-H exchange, the necessary breaking of \ce{H2} into two H atoms
requires more energy than the \ce{H + H2Cl+} process, which already
possesses a barrier. If the process occurs through the formation of
\ce{H4Cl+}, our ab initio calculations show the potential to be
repulsive. Therefore the only way to obtain the observed 3:1 OPR for
\ce{H2Cl+} seems to be via hydrogen abstraction, and not via
scrambling with thermalization afterwards. Of course, it is possible
that there is another and more efficient direction to thermal
equilibrium, which we are missing, but it is, in our view, unlikely.

Thus our tentative conclusion is that under diffuse or translucent
conditions, the measured OPR values for \ce{H2Cl+} are best explained
in terms of a hydrogen abstraction mechanism leading to formation by
mainly the reaction~(\ref{reaction4}) followed by no spin dependence
in the destruction of this ion and, most critically, no likely rapid
interconversion by reaction between the ortho and para forms. But does
reaction~(\ref{reaction4}) indeed proceed via hydrogen
  abstraction?  In order to explore more carefully the different
mechanism possibilities we determined the PES for the system and ran
thousands of QCT to study the dependence on impact parameter and
energy for the fraction of trajectories that {\it (i)} undergo no
reaction, {\it (ii)} undergo an H-abstraction, and {\it (iii)}
undergo an H-exchange. The results are presented in the following
section.

\section{Calculations of the \ce{H2 + HCl+} and \ce{H + H2Cl+} reactions}
\label{sec:theo}

\subsection{Ab initio calculations}

\begin{figure}
\centering
\includegraphics[scale=0.3]{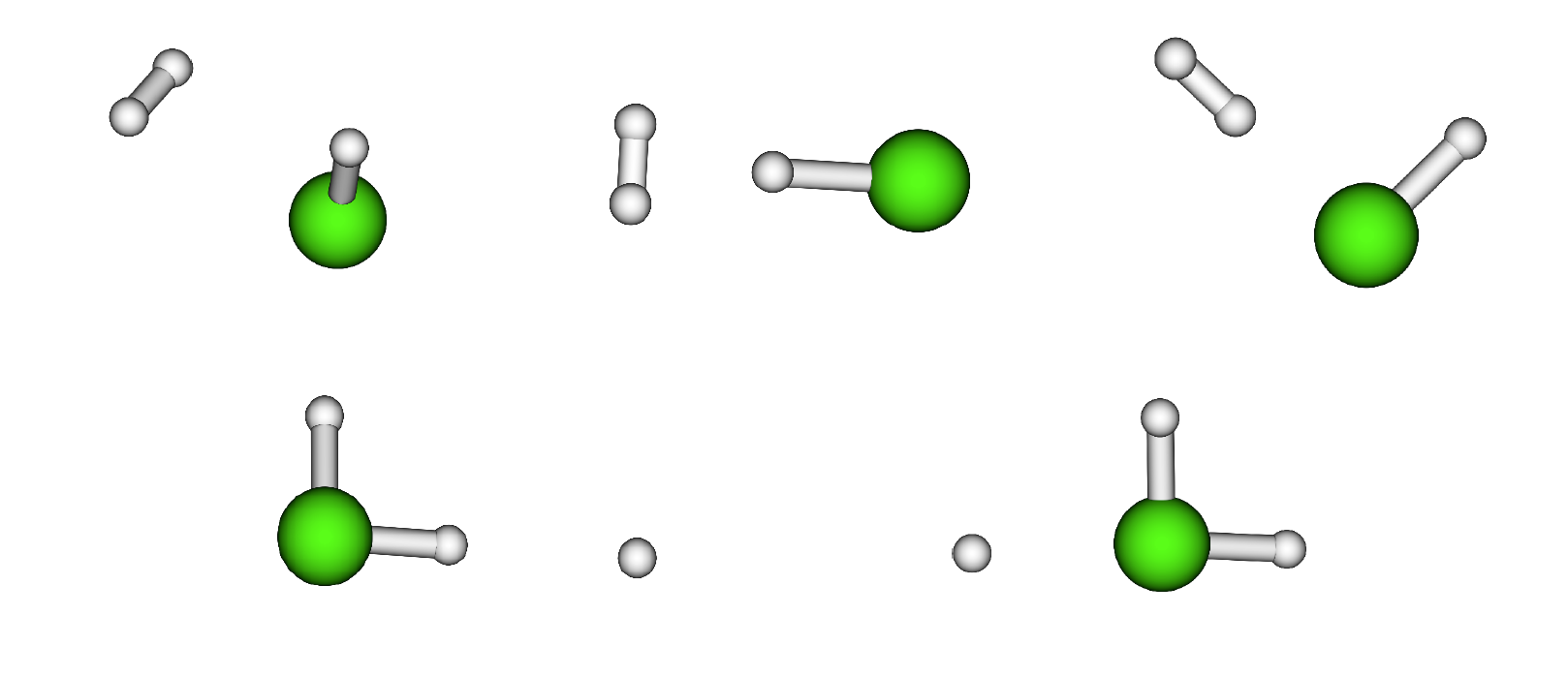}
\caption{Possible approaches of \ce{H2} toward \ce{HCl+} (top) and of H
  toward \ce{H2Cl+} (bottom) considered in this work.}
\label{Fig:theo_1}
\end{figure}

The stationary points on the \ce{HCl+(X^2$\Pi$) + H2(X^1$\Sigma$_g) ->
  H2Cl+(X^1A_1) + H($^2$S)} and \ce{H2Cl+(X^1A1) + H($^2$S) ->
  H($^2$S) + H2Cl+(X^1A_1)} reaction paths were first searched using
the hybrid generalized gradient approximation ( GGA) and B3LYP
functional \citep{stephens1994} in conjunction with the 6-311++G(d,p)
Pople-style triple-zeta-valence basis
\citep{dunning1989,woon1993,kendall1996}. This was followed by
second-order Moller-Plesset perturbation theory (MP2) with the
correlation-consistent polarized valence triple-zeta (cc-pvtz) basis
\citep{gonzalez1989,gonzalez1990}. All possible approaches of \ce{H2}
toward \ce{HCl+} and of H toward \ce{H2Cl+} were considered as shown
in Fig.~\ref{Fig:theo_1}. The character of each structure was
confirmed by a vibrational analysis carried out at the same level
making use of analytical second derivative methods. The assignment of
the saddle points was performed using an intrinsic reaction coordinate
(IRC) calculation. All calculations were carried out by Gaussian 09
software \citep{gaussian09}.  Finally, minima and transition states
were recalculated with the explicitly correlated (F12b) version of the
unrestricted coupled-cluster method with singles, doubles, and
perturbative triples \citep[UCCSD(T)-F12b,][]{knizia2008,knizia2009},
as implemented in MOLPRO \citep{MOLPRO_brief}. The
correlation-consistent polarized valence quadruple-zeta basis set with
F12 optimized \citep[cc-pVQZ-F12,][]{peterson2008} was used.

\begin{figure}
\centering
\includegraphics[scale=0.6]{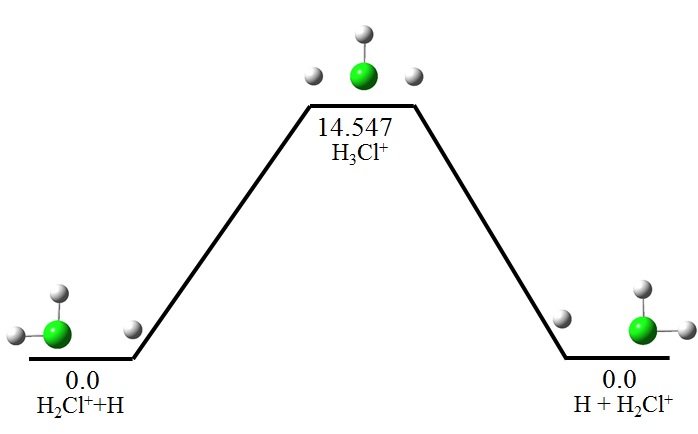}
\caption{UCCSD(T)-F12/ cc-pVQZ-F12 lowest energy profile for the
  \ce{H^'H^{''}Cl+(X^1A_1) + H(^2S) -> H^'(^2S) + H^{''}HCl+(X^1A_1)}
  H-exchange reaction. The energies of the stationary points are given in
  kcal/mol relative to \ce{H + H2Cl+}. Zero point energies are not
  considered.}
\label{Fig:theo_3}
\end{figure}

The resulting lowest energy profiles are shown in
Fig.s~\ref{Fig:theo_2} and \ref{Fig:theo_3}, the geometries of the
stationary structures are given Tables~\ref{tab:theo_1}
and~\ref{tab:theo_2} and the corresponding frequencies in
Tables~\ref{tab:theo_3} and \ref{tab:theo_4}. Figure~\ref{Fig:theo_2}
shows that the formation of \ce{H2Cl+} from \ce{H2} and \ce{HCl+}
involves a barrierless hydrogen abstraction mechanism with the
formation of a stable intermediate complex M3. A proton that
approaches one of the hydrogens of \ce{H2Cl+} leads to the formation
of the M3 complex of Fig.~\ref{Fig:theo_2}. This barrierless approach
corresponds to the reverse path of the \ce{HCl+(X^2$\Pi$) +
  H2(X^1$\Sigma$_g) -> H2Cl+(X^1A_1) + H($^2$S)} reaction (from the
products to M3). On the contrary, the approach of the proton toward
the chlorine of \ce{H2Cl+} involves a very high energy barrier of
14.547 kcal/mol (7320 K), as can be seen in Fig.~\ref{Fig:theo_3}.

From the two reactions studied only the barrierless \ce{HCl+(X^2$\Pi$)
  + H2(X^1$\Sigma$_g) -> H2Cl+(X^1A_1) + H($^2$S)} reaction appears to
be relevant for the question addressed in this study. To quantify its
efficiency, QCT calculations were undertaken. A PES was constructed
from the ab intio points spread in all possible configurations
accessible by these two reactions. These ab initio points were firstly
sampled in the normal mode space of the reactants, products, and
stationary points. Additional sets of geometries were then generated
by running direct dynamics at the MP2/6-31G(d) level with different
initial geometries in the interaction region, using Gaussian~09
\citep{gaussian09}. These points were recalculated at the same
UCCSD(T)-F12b/cc-pVQZ-F12 level sited above. The exothermicity of the
reaction \ce{HCl+ + H2 -> H2Cl+ + H} was computed at the UCCSD(T)-F12b
level with different basis sets cc-pVTZ-F12 and cc-pVQZ-F12, and the
difference is only 0.2\%.  Based on these points, a primitive PES was
first constructed. Classical trajectories with many different initial
conditions starting in the reactant, product, and interaction regions
were run on the PES to explore the configuration space and improve the
PES with new points. These new points were added if they were not too
close to the existing points, which was judged by the Euclidean
distance ($<0.05$~\ang). This procedure was repeated until convergence
was achieved.

\subsection{PES fitting}

A total of 37767 points below 69~kcal/mol relative to the global
minimum on the PES were selected and fit using the permutation
invariant polynomial-neural network (PIP-NN) method
\citep{jiang2013,li2013,jiang2016}. Although no symmetry is considered
in the ab initio calculations, the PIP-NN method enforces the
invariance of the PES when two identical atoms (H in this case) are
exchanged. To enforce the permutation symmetry of the system, symmetry
functions in the form of PIPs were used in the input layer of the NN,
and the permutation symmetry of all three H atoms was considered. In
the PIP-NN approach, the symmetrized polynomials are \citep{xie2010}:

\begin{ceqn}
\begin{align}
  G=\hat S \prod_{i<j}^N p_{ij}^{l_{ij}} \,,
\end{align}
\end{ceqn}
where $l_{ij}$ is the order of the monomial, $\hat S$ is the
symmetrization operator, $N$ is the number of atoms, and
$p_{ij}=\exp(-\alpha r_{ij})$ are the Morse-like variables with
$\alpha=1.0$ \ang$^{-1}$ and the internuclear distances $r_{ij}$ \citep{braams2009}.
All 50 PIPs up to the fourth order were used in the input layer and
the NN architecture was selected to be of 20 and 80 neurons for the
two hidden layers, resulting in 2781 parameters. The NN fittings were
trained using the Levenberg-Marquardt algorithm \citep{hagan1994} and the root mean
square error (RMSE), defined as:

\begin{ceqn}
\begin{align}
  {\rm RMSE}=\sqrt {\sum_{i=1}^{N_{\rm data}}w\left(E_i^{\rm output} -
    E_i^{\rm target}\right)^2/N_{\rm data}} \,,
\end{align}
\end{ceqn}
where $w$ is the weighing function $0.1/(0.1 + E)$ with $E$ relative
to the global minimum, and $E_i^{\rm target}$ and $E_i^{\rm output}$
are the input and fitted energies, respectively. To avoid overfitting,
the data were randomly divided into three sets, namely, the training
(90\% of the data points), validation (5\%), and testing (5\%)
sets. The three best PIP-NN fits have RMSEs for training,
  validation, testing sets and a maximum deviation of 23.0, 43.8, 80.7
  and 10909.9~cal/mol for the first one, 23.0, 57.7, 73.8 and
  9766.1~cal/mol for the second one, and 25.3, 48.3, 73.8 and
  9932.2~cal/mol for the third one. The RMSE of the averaged PIP-NN
PES is 23.0~cal/mol ($\approx 1.0$~meV~$\approx 11.6$~K), and the
maximum deviation is 4~kcal/mol.

\begin{figure}
\centering
\includegraphics[scale=0.65]{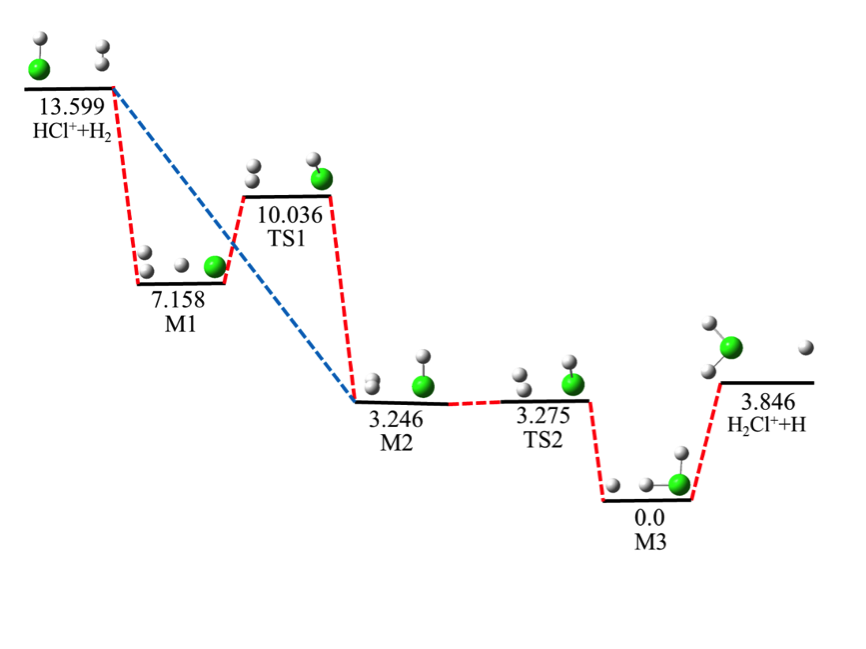}
\caption{Lowest energy UCCSD(T)-F12/ cc-pVQZ-F12 energy profile for the
  \ce{HCl+(X^2$\Pi$) + H2(X^1$\Sigma$_g) -> H2Cl+(X^1A_1) + H($^2$S)}
  reaction. The energies of the stationary points are given in
  kcal/mol relative to the M3 minimum (1 kcal/mol = 503.2 K).  Zero point energies are not
included.}
\label{Fig:theo_2}
\end{figure}

The equilibrium geometries and harmonic frequencies of the stationary
structures of the reaction profile of Fig.~\ref{Fig:theo_2}
(hereafter designed as ab initio values) are well reproduced by the
fitted PES, as shown in Tables~\ref{tab:theo_1} and
\ref{tab:theo_3}. Since the molecular system is very floppy near M2 and
TS2, as shown in Fig.~\ref{Fig:theo_2}, relatively large differences
in the harmonic frequencies of the low-frequency modes exist between
the PES and ab initio values at the M2 and TS2 geometries.

\subsection{QCT calculations}

QCT calculations in this work were implemented in VENUS
\citep{hu1991}. The possible violation of the zero-point energy (ZPE)
is a major deficiency in the QCT method, especially at low
temperatures, since the ZPE then possesses a large fraction of the
total energy. Following our previous work \citep{legal2016} to avoid a
violation of the ZPE, the approach proposed by Hase and coworkers
\citep{paul2016} was used. Since the \ce{HCl+(X^2$\Pi$) +
  H2(X^1$\Sigma$_g) -> H2Cl+(X^1A_1) + H($^2$S)} reaction is
barrierless and exothermic (9.75 kcal/mol), only the violation of the
ZPE in the reactant channel was considered. The calculated reaction
energy with the zero-point energy correction is $\approx
10.25$~kcal/mol, still about 3 kcal/mol smaller than the value based
on the JANAF table of $\approx 13.3$~kcal/mol \citep{lias1988}. The
latter was obtained indirectly and might contain undetermined
uncertainties. The CCSD(T) method used here is considered the
``gold standard'' in computing thermodynamics energies with an
uncertainty that should not exceed a few kcal/mol \citep[][and
references therein]{knizia2009} and is thus fairly
trustworthy. When a nonreactive trajectory exits the strongly
interacting region, the vibrational energies of the \ce{HCl+} and
\ce{H2} are calculated. If the energy is less than the ZPE, the
momenta of all atoms in the system are reversed and the trajectory is
forced back to the strongly interacting region without violating
energy conservation. Only those trajectories with \ce{HCl+}/\ce{H2}
internal energies larger than the ZPEs are accepted.

The trajectories were initiated with a 15.0~\ang\ separation between
reactants, and terminated when products reached a separation of
8.0~\ang. The ro-vibrational states of the reactants \ce{HCl+}/\ce{H2}
and relative translational energies were sampled from the Boltzmann
distribution at a specific temperature. The propagation time step was
chosen as 0.1~fs. Trajectories were discarded if (a) the propagation
time reached 30~ps in each interval of two consecutive momentum
reversing operations; (b) the number of momentum reversing
trajectories exceeds 100; or (c) the total energy failed to converge
to 0.05~kcal/mol. The total reaction probability ($P$) for the
\ce{HCl+(X^2$\Pi$) + H2(X^1$\Sigma$_g)} reaction is

\begin{ceqn}
\begin{align}
  P= \frac{N_{\rm r}}{N_{\rm total}} \,,
\end{align}
\end{ceqn}
where $N_{\rm r}$ and $N_{\rm total}$ are the number of reactive trajectories
and total number of trajectories, respectively. The standard error is
given by: 
\begin{ceqn}
\begin{align}
  \Delta= \sqrt{(N_{\rm total}-N_{\rm r})/N_{\rm total}N_{\rm r}} \,.
\end{align}
\end{ceqn}

\begin{figure}
\centering
\includegraphics[scale=0.7]{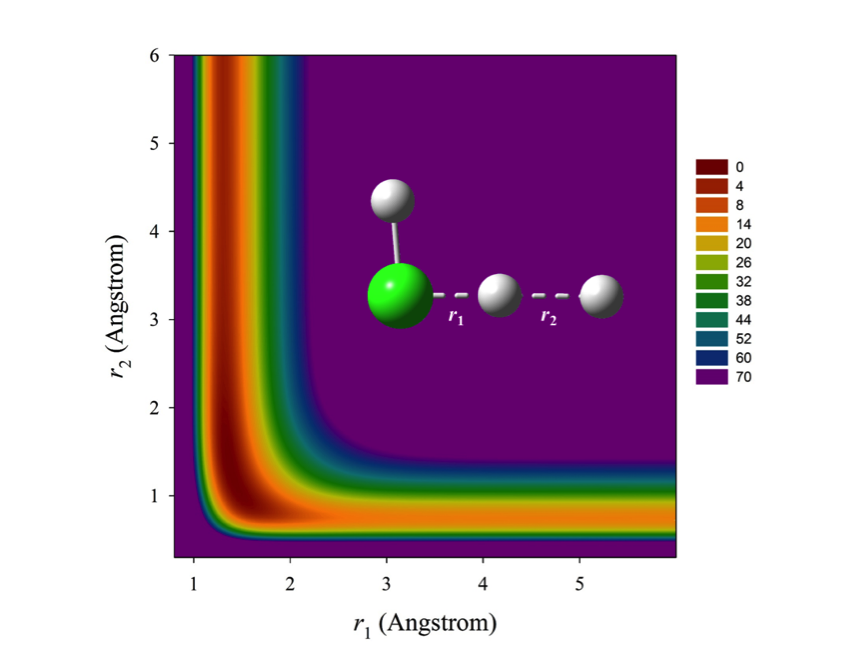}
\caption{Contour plot for the H-atom abstraction reaction
  process \ce{HCl+(X^2$\Pi$) + H2(X^1$\Sigma$_g) -> H2Cl+(X^1A_1) +
    H($^2$S)} reaction along the two reaction coordinates, \ce{Cl-H^'}
  ($r_1$) and \ce{H^'-H^{"}} ($r_2$) distances, with all the other
  internal coordinates optimized. The energies are in kcal/mol
  relative to the global minimum M3.}
\label{Fig:theo_4}
\end{figure}

In the QCT calculations, more than 20000 trajectories were calculated
at each temperature, and the statistical errors of the total reaction
probabilities are all small. The \ce{HCl+}(X$^2$$\Pi$) +
\ce{H2}(X$^1$$\Sigma_{\rm g}$) $\rightarrow$ \ce{H2Cl+}(X$^1$A$_1$) + H($^2$S)
reaction has three typical processes, the first is the nonreactive
channel, which has three possible outcomes starting from HCl$^+$ +
H'H": HCl$^+$ + H'H", H'Cl$^+$ + HH", and H"Cl$^+$ + HH'.  It should
be noted that the only nonreactive trajectories found in our QCT
calculations are those in which no exchange of H atoms take place;
i.e., HCl$^{+}$ + H'H" $\rightarrow$ HCl$^{+}$ + H'H".  The second
process is a reactive H-abstraction channel: HCl$^+$ + H'H"
$\rightarrow$ HH'Cl$^+$ + H"/ HH"Cl$^+$ + H', an example of which is
shown in Fig.~\ref{Fig:theo_4}. The last one is a reactive H-exchange
channel: HCl$^+$ + H'H" $\rightarrow$ H'H"Cl$^{+ }$+ H. The numbers of
these three types of trajectories at each temperature are listed in
Table~\ref{tab:theo_5}.  It can be readily seen from
Table~\ref{tab:theo_5} that the H-abstraction products are dominant
for all the studied temperatures ranging from 20 to 300~K. As the
temperature decreases, the percentage of the H-exchange trajectories
increases.  However, at the lowest temperature studied, 20~K, the
ratio between the H-exchange and H-abstraction channels is still small
(0.85\%).

The potential and trajectory studies lead to two principal
conclusions: (a) the \ce{HCl+(X^2$\Pi$) + H2(X^1$\Sigma$_g)} reaction
proceeds overwhelmingly by an H-abstraction mechanism, which
leads to a 3:1 OPR, and (b) the nascent OPR ratio cannot be altered by
thermalization reactions involving atomic hydrogen because the process
has a large barrier.

\section{Summary and conclusions} 
\label{sec:conclu}

We have presented a detailed study of the nuclear-spin chemistry of
\ce{H2Cl+} in which we show that one reaction - HCl$^{+}$ + H$_{2}$
$\rightarrow$ H$_{2}$Cl$^{+}$ + H - is mainly responsible for the
production of this ion and the OPR obtained during its formation under
diffuse or translucent conditions. We then study this reaction in some
detail using a QCT method based on the ab inito PES. This research has
been undertaken in the light of new ALMA observations of both
  ortho- and para-\ce{H2Cl+} forms in the z = 0.89 absorber toward the
  quasar \PKS1830, along two independent sight lines with
  different physical properties. In both sightlines, we measured an
  OPR consistent with a 3:1 ratio. As explained in the present study,
for molecules containing two hydrogen atoms and presenting specific
symmetry properties, such an OPR is consistent with the spin
statistical weight, which can be obtained by high temperatures in
which many rotational levels are populated by thermalization
processes, or by some mechanism involving reactions that drive the OPR
formation, which is subsequently not thermalized. Indeed, two extreme
non-thermal reaction mechanisms can be distinguished: {\it (i)} a
simple hydrogen atom abstraction process or {\it (ii)} a full
scrambling process in which both H-abstraction and H-exchange can
occur, as described in Sect.~\ref{sec:chem}. A simple H-abstraction
can naturally lead to an OPR of 3:1, as is the case here, while
scrambling will lead to different values depending upon the OPR of
\ce{H2}, known to trace the thermal history of its environment. In
other words, the resulting so-called ``formation OPR'', in which
destruction and thermalization are not yet considered, depends upon
whether the reaction mechanism has a potential well so that {\it (i)}
an intermediate complex can be formed and {\it (ii)} its lifetime is
long enough to allow the reactions to occur through a scrambling
process \citep{oka2004,herbst2015}.

As a consequence, the theory of \citet{oka2004}, which is used as
standard, in which scrambling of H atoms via a long-lived complex is
most frequently assumed, should be used with care. It should not be
used unambiguously without detailed knowledge of the dynamics of a
chemical reaction, which requires the study of the PES as well as
trajectory studies to determine the dominant mechanism or mechanisms
leading to products. The theory of \citet{oka2004} also requires
exothermic reactions where many product rotational states are
populated. If this is not the case, a 3:1 value is unlikely to be
achieved. For example if only rotational levels with quantum numbers
$J$ = 0 (para) and $J$=1 (ortho) can be produced, there are nine times
as many ortho states as para states. These different types of
mechanisms have been explored to interpret the synthesis of the OPR of
\ce{H3+} \citep{crabtree2011b,crabtree2011c} and subsequently studied
in more detail for the specific system \ce{H2 + H3+}
\citep{gomez-carrasco2012}. More recently, this distinction of
mechanisms has been revived in order to explain the observed OPRs of
\ce{H2O+} \citep{herbst2015} and \ce{H2Cl+} \citep{neufeld2015}.

Our present study involves the use of QCT calculations on a highly
accurate global PES. In this study we show that despite a potential
minimum, the mechanism of the reaction between H$_{2}$ and HCl$^{+}$
goes overwhelmingly by a simple H-abstraction angular momentum
considerations to a 3:1 OPR. This is consistent with all observations
of the chloronium ion including the new observations discussed here.
However, one must also consider how the OPR is affected by destruction
and thermalization reactions.  Although destruction processes are
unlikely to affect the OPR of H$_{2}$Cl$^{+}$, thermalization by
reactions with reactive species might be possible, even though such
a process with atomic hydrogen has been shown to have a high
barrier in the present study (Section~\ref{sec:theo}) and are even
less plausible with molecular hydrogen as discussed in Sect.
\ref{subsec:modeling}. For other molecules, given enough time,
thermalization might lead to a thermal value of OPR which could
reflect the various other temperatures in the sources, as found for
instance for the \ce{NH2} case \citep{legal2016}. From these examples,
one must be more careful in assuming that scrambling always takes
place when potential minima leading to reaction complexes are
present. Although trajectory calculations for reactions involving
large numbers of atoms can prove tedious, one can learn a considerable
amount from a calculation of the mimum energy pathway.

\begin{acknowledgement}
  The authors thank the anonymous referee for nice and interesting
  comments. RL is very grateful to Valentine Wakelam for allowing her
  to use Nautilus in independent research. The authors also warmly
  thank John Black and Evelyne Roueff for very interesting comments
  which helped to improve the present manuscript. EH wishes to thank
  the National Science Foundation for continuing to support the
  astrochemistry program at the University of Virginia. This paper
  makes use of the following ALMA data: \#2013.1.00020.S. ALMA is a
  partnership of ESO (representing its member states), NSF (USA) and
  NINS (Japan), together with NRC (Canada) and NSC and ASIAA (Taiwan)
  and KASI (Republic of Korea), in cooperation with the Republic of
  Chile. The Joint ALMA Observatory is operated by ESO, AUI/NRAO and
  NAOJ. CX and HG are supported by US Department of Energy (Grant
  No. DE-SC0015997). The computations were performed at the Center for
  Advanced Research Computing (CARC) at UNM. CX thanks Jun Li for many
  useful discussions on PIP-NN method. DT acknowledges the HPC
  resources from GENCI-[CCRT/CINES/IDRIS] (grants 2016 and 2017
  [x2016085116-A0020805116] for computing time as well as the
  Programme National “Physique et Chimie du Milieu Interstellaire”
  (PCMI) of CNRS/INSU with INC/INP co-funded by CEA and CNES for its
  support.

\end{acknowledgement}

\bibliographystyle{aa}
\bibliography{biblio-H2Clp}

\begin{thebibliography}{71}
\expandafter\ifx\csname natexlab\endcsname\relax\def\natexlab#1{#1}\fi

\bibitem[{Braams \& Bowman(2009)}]{braams2009}
Braams, B.~J. \& Bowman, J.~M. 2009, Int. Rev. Phys. Chem., 28, 577

\bibitem[{Crabtree {et~al.}(2011{\natexlab{a}})Crabtree, {Indriolo}, {Kreckel},
  {Tom}, \& {McCall}}]{crabtree2011a}
Crabtree, K.~N., {Indriolo}, N., {Kreckel}, H., {Tom}, B.~A., \& {McCall},
  B.~J. 2011{\natexlab{a}}, \apj, 729, 15


\bibitem[{Crabtree {et~al.}(2011{\natexlab{b}})Crabtree, Tom, \&
  McCall}]{crabtree2011b}
Crabtree, K.~N., Tom, B.~A., \& McCall, B.~J. 2011{\natexlab{b}}, J. Chem.
  Phys., 134, 194310

\bibitem[{Crabtree {et~al.}(2011{\natexlab{c}})Crabtree, Kauffman, Tom,
  Be√ßka, McGuire, \& McCall}]{crabtree2011c}
Crabtree, K.~N., Kauffman, C.~A., Tom, B.~A., {et~al.} 2011{\natexlab{c}}, J.
  Chem. Phys., 134, 194311


\bibitem[{{Crockett} {et~al.}(2014){Crockett}, {Bergin}, {Neill}, {Black},
  {Blake}, \& {Kleshcheva}}]{crockett2014}
{Crockett}, N.~R., {Bergin}, E.~A., {Neill}, J.~L., {et~al.} 2014, \apj, 781,
  114

\bibitem[{{Dalgarno} {et~al.}(1974){Dalgarno}, {de Jong}, {Oppenheimer}, \&
  {Black}}]{dalgarno1974}
{Dalgarno}, A., {de Jong}, T., {Oppenheimer}, M., \& {Black}, J.~H. 1974,
  \apjl, 192, L37

\bibitem[{{de Ruette} {et~al.}(2016){de Ruette}, {Miller}, {O'Connor},
  {Urbain}, {Buzard}, {Vissapragada}, \& {Savin}}]{deRuette2016}
{de Ruette}, N., {Miller}, K.~A., {O'Connor}, A.~P., {et~al.} 2016, \apj, 816,
  31

\bibitem[{{Dunning}(1989)}]{dunning1989}
{Dunning}, T.~H. 1989, J. Chem. Phys., 90, 1007

\bibitem[{{Faure} {et~al.}(2013){Faure}, {Hily-Blant}, {Le Gal}, {Rist}, \&
  {Pineau des For{\^e}ts}}]{faure2013}
{Faure}, A., {Hily-Blant}, P., {Le Gal}, R., {Rist}, C., \& {Pineau des
  For{\^e}ts}, G. 2013, \apjl, 770, L2

\bibitem[{{Flagey} {et~al.}(2013){Flagey}, {Goldsmith}, {Lis}, {Gerin},
  {Neufeld}, {Sonnentrucker}, {De Luca}, {Godard}, {Goicoechea}, {Monje}, \&
  {Phillips}}]{flagey2013}
{Flagey}, N., {Goldsmith}, P.~F., {Lis}, D.~C., {et~al.} 2013, \apj, 762, 11

\bibitem[{Frisch {et~al.}(2009)Frisch, Trucks, Schlegel, Scuseria, Robb,
  Cheeseman, Scalmani, Barone, Mennucci, Petersson, Nakatsuji, Caricato, Li,
  Hratchian, Izmaylov, Bloino, Zheng, Sonnenberg, Hada, Ehara, Toyota, Fukuda,
  Hasegawa, Ishida, Nakajima, Honda, Kitao, Nakai, Vreven, Montgomery, Peralta,
  Ogliaro, Bearpark, Heyd, Brothers, Kudin, Staroverov, Kobayashi, Normand,
  Raghavachari, Rendell, Burant, Iyengar, Tomasi, Cossi, Rega, Millam, Klene,
  Knox, Cross, Bakken, Adamo, Jaramillo, Gomperts, Stratmann, Yazyev, Austin,
  Cammi, Pomelli, Ochterski, Martin, Morokuma, Zakrzewski, Voth, Salvador,
  Dannenberg, Dapprich, Daniels, Farkas, Foresman, Ortiz, Cioslowski, \&
  Fox}]{gaussian09}
Frisch, M.~J., Trucks, G.~W., Schlegel, H.~B., {et~al.} 2009, {Gaussian 09,
  Revision B.01, Gaussian, Inc., Wallingford CT}

\bibitem[{{Gerin} {et~al.}(2013){Gerin}, {de Luca}, {Lis}, {Kramer}, {Navarro},
  {Neufeld}, {Indriolo}, {Godard}, {Le Petit}, {Peng}, {Phillips}, \&
  {Roueff}}]{gerin2013}
{Gerin}, M., {de Luca}, M., {Lis}, D.~C., {et~al.} 2013, J. Phys. Chem. A, 117,
  10018

\bibitem[{{Gerin} {et~al.}(2016){Gerin}, {Neufeld}, \&
  {Goicoechea}}]{gerin2016}
{Gerin}, M., {Neufeld}, D.~A., \& {Goicoechea}, J.~R. 2016, \araa, 54, 181

\bibitem[{{Glassgold} \& {Langer}(1974)}]{glassgold1974}
{Glassgold}, A.~E. \& {Langer}, W.~D. 1974, \apj, 193, 73

\bibitem[{Gomez-Carrasco {et~al.}(2012)Gomez-Carrasco, Gonzalez-Sanchez,
  Aguado, Sanz-Sanz, Zanchet, \& Roncero}]{gomez-carrasco2012}
Gomez-Carrasco, S., Gonzalez-Sanchez, L., Aguado, A., {et~al.} 2012, J. Chem.
  Phys., 137, 094303

\bibitem[{Gonzalez \& Schlegel(1989)}]{gonzalez1989}
Gonzalez, C. \& Schlegel, H.~B. 1989, J. Chem. Phys., 90, 2154

\bibitem[{Gonzalez \& Schlegel(1990)}]{gonzalez1990}
Gonzalez, C. \& Schlegel, H.~B. 1990, J. Phys. Chem., 94, 5523

\bibitem[{Hagan \& Menhaj(1994)}]{hagan1994}
Hagan, M.~T. \& Menhaj, M.~B. 1994, IEEE Transactions on Neural Networks, 5,
  989

\bibitem[{{Henkel} {et~al.}(2008){Henkel}, {Braatz}, {Menten}, \&
  {Ott}}]{henkel2008}
{Henkel}, C., {Braatz}, J.~A., {Menten}, K.~M., \& {Ott}, J. 2008, \aap, 485,
  451

\bibitem[{{Henkel} {et~al.}(2009){Henkel}, {Menten}, {Murphy}, {Jethava},
  {Flambaum}, {Braatz}, {Muller}, {Ott}, \& {Mao}}]{henkel2009}
{Henkel}, C., {Menten}, K.~M., {Murphy}, M.~T., {et~al.} 2009, \aap, 500, 725

\bibitem[{Herbst(2015)}]{herbst2015}
Herbst, E. 2015, EPJ Web of Conferences, 84, 06002

\bibitem[{Hu {et~al.}(1991)Hu, Hase, \& Pirraglia}]{hu1991}
Hu, X., Hase, W.~L., \& Pirraglia, T. 1991, J. Comput. Chem., 12, 1014

\bibitem[{{Jauncey} {et~al.}(1991){Jauncey}, {Reynolds}, {Tzioumis}, {Muxlow},
  {Perley}, {Murphy}, {Preston}, {King}, {Patnaik}, {Jones}, {Meier}, {Bird},
  {Blair}, {Bunton}, {Clay}, {Costa}, {Duncan}, {Ferris}, {Gough}, {Hamilton},
  {Hoard}, {Kemball}, {Kesteven}, {Lobdell}, {Luiten}, {Mcculloch}, {Murray},
  {Nicholson}, {Rao}, {Savage}, {Sinclair}, {Skjerve}, {Taaffe}, {Wark}, \&
  {White}}]{jauncey1991}
{Jauncey}, D.~L., {Reynolds}, J.~E., {Tzioumis}, A.~K., {et~al.} 1991, \nat,
  352, 132

\bibitem[{{Jiang} \& {Guo}(2013)}]{jiang2013}
{Jiang}, B. \& {Guo}, H. 2013, \jcp, 139, 054112

\bibitem[{Jiang {et~al.}(2016)Jiang, Li, \& Guo}]{jiang2016}
Jiang, B., Li, J., \& Guo, H. 2016, Int. Rev. Phys. Chem., 35, 479

\bibitem[{{Jura}(1974)}]{jura1974}
{Jura}, M. 1974, \apjl, 190, L33

\bibitem[{Kendall {et~al.}(1992)Kendall, Jr., \& Harrison}]{kendall1996}
Kendall, R.~A., Jr., T. H.~D., \& Harrison, R.~J. 1992, J. Chem. Phys., 96,
  6796

\bibitem[{{Knizia} {et~al.}(2009){Knizia}, {Adler}, \& {Werner}}]{knizia2009}
{Knizia}, G., {Adler}, T.~B., \& {Werner}, H.-J. 2009, \jcp, 130, 054104

\bibitem[{{Knizia} \& {Werner}(2008)}]{knizia2008}
{Knizia}, G. \& {Werner}, H.-J. 2008, \jcp, 128, 154103

\bibitem[{{Koopmans} \& {de Bruyn}(2005)}]{koopmans2005}
{Koopmans}, L.~V.~E. \& {de Bruyn}, A.~G. 2005, \mnras, 360, L6

\bibitem[{{Le Gal} {et~al.}(2016){Le Gal}, {Herbst}, {Xie}, {Li}, \&
  {Guo}}]{legal2016}
{Le Gal}, R., {Herbst}, E., {Xie}, C., {Li}, A., \& {Guo}, H. 2016, \aap, 596,
  A35

\bibitem[{{Le Gal} {et~al.}(2014{\natexlab{a}}){Le Gal}, {Hily-Blant}, {Faure},
  {Pineau des For{\^e}ts}, {Rist}, \& {Maret}}]{legal2014a}
{Le Gal}, R., {Hily-Blant}, P., {Faure}, A., {et~al.} 2014{\natexlab{a}}, \aap,
  562, A83

\bibitem[{{Le Gal} {et~al.}(2014{\natexlab{b}}){Le Gal}, {Hily-Blant}, \&
  {Faure}}]{legal2014b}
{Le Gal}, R., {Hily-Blant}, P., \& {Faure}, A. 2014{\natexlab{b}}, in
  SF2A-2014: Proceedings of the Annual meeting of the French Society of
  Astronomy and Astrophysics, ed. J.~{Ballet}, F.~{Martins}, F.~{Bournaud},
  R.~{Monier}, \& C.~{Reyl{\'e}}, 397--401



\bibitem[{Le~Petit {et~al.}(2006)Le~Petit, Nehm\'e, Le~Bourlot, \&
  Roueff}]{lepetit2006}
Le~Petit, F., Nehm\'e, C., Le~Bourlot, J., \& Roueff, E. 2006, \apjs, 164, 506

\bibitem[{{Li} {et~al.}(2013){Li}, {Jiang}, \& {Guo}}]{li2013}
{Li}, J., {Jiang}, B., \& {Guo}, H. 2013, \jcp, 139, 204103

\bibitem[{{Lias} {et~al.}(1988){Lias}, {Bartmess}, {Liebman}, {Holmes},
  {Levin}, \& {Mallard}}]{lias1988}
{Lias}, S.~G., {Bartmess}, J.~E., {Liebman}, J.~F., {et~al.} 1988, J. Phys.
  Chem. Ref. Data, Suppl. 1, 17

\bibitem[{{Lique} {et~al.}(2014){Lique}, {Honvault}, \& {Faure}}]{lique2014}
{Lique}, F., {Honvault}, P., \& {Faure}, A. 2014, ArXiv e-prints
  [\eprint[arXiv]{1402.5292}]

\bibitem[{{Lis} {et~al.}(2013){Lis}, {Bergin}, {Schilke}, \& {van
  Dishoeck}}]{lis2013}
{Lis}, D.~C., {Bergin}, E.~A., {Schilke}, P., \& {van Dishoeck}, E.~F. 2013, J.
  Phys. Chem. A, 117, 9661

\bibitem[{{Lis} {et~al.}(2010){Lis}, {Pearson}, {Neufeld}, {Schilke},
  {M{\"u}ller}, {Gupta}, {Bell}, {Comito}, {Phillips}, {Bergin}, {Ceccarelli},
  {Goldsmith}, {Blake}, {Bacmann}, {Baudry}, {Benedettini}, {Benz}, {Black},
  {Boogert}, {Bottinelli}, {Cabrit}, {Caselli}, {Castets}, {Caux},
  {Cernicharo}, {Codella}, {Coutens}, {Crimier}, {Crockett}, {Daniel}, {Demyk},
  {Dominic}, {Dubernet}, {Emprechtinger}, {Encrenaz}, {Falgarone}, {Fuente},
  {Gerin}, {Giesen}, {Goicoechea}, {Helmich}, {Hennebelle}, {Henning},
  {Herbst}, {Hily-Blant}, {Hjalmarson}, {Hollenbach}, {Jack}, {Joblin},
  {Johnstone}, {Kahane}, {Kama}, {Kaufman}, {Klotz}, {Langer}, {Larsson}, {Le
  Bourlot}, {Lefloch}, {Le Petit}, {Li}, {Liseau}, {Lord}, {Lorenzani},
  {Maret}, {Martin}, {Melnick}, {Menten}, {Morris}, {Murphy}, {Nagy}, {Nisini},
  {Ossenkopf}, {Pacheco}, {Pagani}, {Parise}, {P{\'e}rault}, {Plume}, {Qin},
  {Roueff}, {Salez}, {Sandqvist}, {Saraceno}, {Schlemmer}, {Schuster}, {Snell},
  {Stutzki}, {Tielens}, {Trappe}, {van der Tak}, {van der Wiel}, {van
  Dishoeck}, {Vastel}, {Viti}, {Wakelam}, {Walters}, {Wang}, {Wyrowski},
  {Yorke}, {Yu}, {Zmuidzinas}, {Delorme}, {Desbat}, {G{\"u}sten}, {Krieg}, \&
  {Delforge}}]{lis2010}
{Lis}, D.~C., {Pearson}, J.~C., {Neufeld}, D.~A., {et~al.} 2010, \aap, 521, L9

\bibitem[{{Mart{\'{\i}}-Vidal} {et~al.}(2014){Mart{\'{\i}}-Vidal}, {Vlemmings},
  {Muller}, \& {Casey}}]{marti-vidal2014}
{Mart{\'{\i}}-Vidal}, I., {Vlemmings}, W.~H.~T., {Muller}, S., \& {Casey}, S.
  2014, \aap, 563, A136

\bibitem[{{Mathis} {et~al.}(1983){Mathis}, {Mezger}, \& {Panagia}}]{mathis1983}
{Mathis}, J.~S., {Mezger}, P.~G., \& {Panagia}, N. 1983, \aap, 128, 212

\bibitem[{{M{\"u}ller} {et~al.}(2015){M{\"u}ller}, {Muller}, {Schilke},
  {Bergin}, {Black}, {Gerin}, {Lis}, {Neufeld}, \& {Suri}}]{muller2015}
{M{\"u}ller}, H.~S.~P., {Muller}, S., {Schilke}, P., {et~al.} 2015, \aap, 582,
  L4

\bibitem[{{Muller} {et~al.}(2013){Muller}, {Beelen}, {Black}, {Curran},
  {Horellou}, {Aalto}, {Combes}, {Gu{\'e}lin}, \& {Henkel}}]{muller2013}
{Muller}, S., {Beelen}, A., {Black}, J.~H., {et~al.} 2013, \aap, 551, A109

\bibitem[{{Muller} {et~al.}(2011){Muller}, {Beelen}, {Gu{\'e}lin}, {Aalto},
  {Black}, {Combes}, {Curran}, {Theule}, \& {Longmore}}]{muller2011}
{Muller}, S., {Beelen}, A., {Gu{\'e}lin}, M., {et~al.} 2011, \aap, 535, A103

\bibitem[{{Muller} {et~al.}(2014{\natexlab{a}}){Muller}, {Black}, {Gu{\'e}lin},
  {Henkel}, {Combes}, {G{\'e}rin}, {Aalto}, {Beelen}, {Darling}, {Horellou},
  {Mart{\'{\i}}n}, {Menten}, {V-Trung}, \& {Zwaan}}]{muller2014b}
{Muller}, S., {Black}, J.~H., {Gu{\'e}lin}, M., {et~al.} 2014{\natexlab{a}},
  \aap, 566, L6

\bibitem[{{Muller} {et~al.}(2014{\natexlab{b}}){Muller}, {Combes},
  {Gu{\'e}lin}, {G{\'e}rin}, {Aalto}, {Beelen}, {Black}, {Curran}, {Darling},
  {V-Trung}, {Garc{\'{\i}}a-Burillo}, {Henkel}, {Horellou}, {Mart{\'{\i}}n},
  {Mart{\'{\i}}-Vidal}, {Menten}, {Murphy}, {Ott}, {Wiklind}, \&
  {Zwaan}}]{muller2014a}
{Muller}, S., {Combes}, F., {Gu{\'e}lin}, M., {et~al.} 2014{\natexlab{b}},
  \aap, 566, A112


\bibitem[{{Muller} {et~al.}(2016{\natexlab{a}}){Muller}, {M{\"u}ller}, {Black},
  {Beelen}, {Combes}, {Curran}, {G{\'e}rin}, {Gu{\'e}lin}, {Henkel},
  {Mart{\'{\i}}n}, {Aalto}, {Falgarone}, {Menten}, {Schilke}, {Wiklind}, \&
  {Zwaan}}]{muller2016a}
{Muller}, S., {M{\"u}ller}, H.~S.~P., {Black}, J.~H., {et~al.}
  2016{\natexlab{a}}, \aap, 595, A128

\bibitem[{{Muller} {et~al.}(2016{\natexlab{b}}){Muller}, {Kawaguchi}, {Black},
  \& {Amano}}]{muller2016b}
{Muller}, S., {Kawaguchi}, K., {Black}, J.~H., \& {Amano}, T.
  2016{\natexlab{b}}, \aap, 589, L5



\bibitem[{{Mumma} {et~al.}(1987){Mumma}, {Weaver}, \& {Larson}}]{mumma1987}
{Mumma}, M.~J., {Weaver}, H.~A., \& {Larson}, H.~P. 1987, \aap, 187, 419

\bibitem[{{Neufeld} {et~al.}(2015){Neufeld}, {Black}, {Gerin}, {Goicoechea},
  {Goldsmith}, {Gry}, {Gupta}, {Herbst}, {Indriolo}, {Lis}, {Menten}, {Monje},
  {Mookerjea}, {Persson}, {Schilke}, {Sonnentrucker}, \&
  {Wolfire}}]{neufeld2015}
{Neufeld}, D.~A., {Black}, J.~H., {Gerin}, M., {et~al.} 2015, \apj, 807, 54

\bibitem[{{Neufeld} {et~al.}(2012){Neufeld}, {Roueff}, {Snell}, {Lis}, {Benz},
  {Bruderer}, {Black}, {De Luca}, {Gerin}, {Goldsmith}, {Gupta}, {Indriolo},
  {Le Bourlot}, {Le Petit}, {Larsson}, {Melnick}, {Menten}, {Monje}, {Nagy},
  {Phillips}, {Sandqvist}, {Sonnentrucker}, {van der Tak}, \&
  {Wolfire}}]{neufeld2012}
{Neufeld}, D.~A., {Roueff}, E., {Snell}, R.~L., {et~al.} 2012, \apj, 748, 37

\bibitem[{{Neufeld} \& {Wolfire}(2009)}]{neufeld2009}
{Neufeld}, D.~A. \& {Wolfire}, M.~G. 2009, \apj, 706, 1594

\bibitem[{{Oka}(2004)}]{oka2004}
{Oka}, T. 2004, J. Mol. Spec., 228, 635

\bibitem[{{Pachucki} \& {Komasa}(2008)}]{pachucki2008}
{Pachucki}, K. \& {Komasa}, J. 2008, \pra, 77, 030501

\bibitem[{Paul \& Hase(2016)}]{paul2016}
Paul, A.~K. \& Hase, W.~L. 2016, J. Phys. Chem. A, 120, 372

\bibitem[{{Persson} {et~al.}(2012){Persson}, {De Luca}, {Mookerjea},
  {Olofsson}, {Black}, {Gerin}, {Herbst}, {Bell}, {Coutens}, {Godard},
  {Goicoechea}, {Hassel}, {Hily-Blant}, {Menten}, {M{\"u}ller}, {Pearson}, \&
  {Yu}}]{persson2012}
{Persson}, C.~M., {De Luca}, M., {Mookerjea}, B., {et~al.} 2012, \aap, 543,
  A145

\bibitem[{{Persson} {et~al.}(2016){Persson}, {Olofsson}, {Le Gal},
  {Wirstr{\"o}m}, {Hassel}, {Herbst}, {Olberg}, {Faure}, {Hily-Blant}, {Black},
  {Gerin}, {Lis}, \& {Wyrowski}}]{persson2016}
{Persson}, C.~M., {Olofsson}, A.~O.~H., {Le Gal}, R., {et~al.} 2016, \aap, 586,
  A128

\bibitem[{{Peterson} {et~al.}(2008){Peterson}, {Adler}, \&
  {Werner}}]{peterson2008}
{Peterson}, K.~A., {Adler}, T.~B., \& {Werner}, H.-J. 2008, \jcp, 128, 084102

\bibitem[{{Raich} \& {Good}(1964)}]{raich1964}
{Raich}, J.~C. \& {Good}, Jr., R.~H. 1964, \apj, 139, 1004

\bibitem[{{Ruaud} {et~al.}(2016){Ruaud}, {Wakelam}, \& {Hersant}}]{ruaud2016}
{Ruaud}, M., {Wakelam}, V., \& {Hersant}, F. 2016, \mnras, 459, 3756

\bibitem[{{Schilke} {et~al.}(2010){Schilke}, {Comito}, {M{\"u}ller}, {Bergin},
  {Herbst}, {Lis}, {Neufeld}, {Phillips}, {Bell}, {Blake}, {Cabrit}, {Caux},
  {Ceccarelli}, {Cernicharo}, {Crockett}, {Daniel}, {Dubernet},
  {Emprechtinger}, {Encrenaz}, {Falgarone}, {Gerin}, {Giesen}, {Goicoechea},
  {Goldsmith}, {Gupta}, {Joblin}, {Johnstone}, {Langer}, {Latter}, {Lord},
  {Maret}, {Martin}, {Melnick}, {Menten}, {Morris}, {Murphy}, {Ossenkopf},
  {Pagani}, {Pearson}, {P{\'e}rault}, {Plume}, {Qin}, {Salez}, {Schlemmer},
  {Stutzki}, {Trappe}, {van der Tak}, {Vastel}, {Wang}, {Yorke}, {Yu},
  {Erickson}, {Maiwald}, {Kooi}, {Karpov}, {Zmuidzinas}, {Boogert}, {Schieder},
  \& {Zaal}}]{schilke2010}
{Schilke}, P., {Comito}, C., {M{\"u}ller}, H.~S.~P., {et~al.} 2010, \aap, 521,
  L11

\bibitem[{{Schulz} {et~al.}(2015){Schulz}, {Henkel}, {Menten}, {Muller},
  {Muders}, {Bagdonaite}, \& {Ubachs}}]{schulz2015}
{Schulz}, A., {Henkel}, C., {Menten}, K.~M., {et~al.} 2015, \aap, 574, A108

\bibitem[{Stephens {et~al.}(1994)Stephens, Devlin, Chabalowski, \&
  Frisch}]{stephens1994}
Stephens, P., Devlin, F., Chabalowski, C., \& Frisch, M.~J. 1994, J. Phys.
  Chem., 98, 11623

\bibitem[{{Talbi} {et~al.}(1991){Talbi}, {Defrees}, {Egolf}, \&
  {Herbst}}]{talbi1991}
{Talbi}, D., {Defrees}, D.~J., {Egolf}, D.~A., \& {Herbst}, E. 1991, \apj, 374,
  390

\bibitem[{{Vastel} {et~al.}(2015){Vastel}, {Yamamoto}, {Lefloch}, \&
  {Bachiller}}]{vastel2015}
{Vastel}, C., {Yamamoto}, S., {Lefloch}, B., \& {Bachiller}, R. 2015, \aap,
  582, L3

\bibitem[{{Vissapragada} {et~al.}(2016){Vissapragada}, {Buzard}, {Miller},
  {O'Connor}, {de Ruette}, {Urbain}, \& {Savin}}]{vissapragada2016}
{Vissapragada}, S., {Buzard}, C.~F., {Miller}, K.~A., {et~al.} 2016, \apj, 832,
  31

\bibitem[{Werner {et~al.}(2015)Werner, Knowles, Knizia, Manby, {Sch\"{u}tz},
  {et~al.}}]{MOLPRO_brief}
Werner, H.-J., Knowles, P.~J., Knizia, G., {et~al.} 2015, MOLPRO, version
  2015.1, a package of ab initio programs, see http://www.molpro.net

\bibitem[{{Wiklind} \& {Combes}(1998)}]{wiklind1998}
{Wiklind}, T. \& {Combes}, F. 1998, \apj, 500, 129

\bibitem[{{Winn} {et~al.}(2002){Winn}, {Kochanek}, {McLeod}, {Falco}, {Impey},
  \& {Rix}}]{winn2002}
{Winn}, J.~N., {Kochanek}, C.~S., {McLeod}, B.~A., {et~al.} 2002, \apj, 575,
  103

\bibitem[{{Woon} \& {Dunning}(1993)}]{woon1993}
{Woon}, D.~E. \& {Dunning}, T.~H. 1993, J. Chem. Phys., 98, 1358

\bibitem[{Xie \& Bowman(2010)}]{xie2010}
Xie, Z. \& Bowman, J.~M. 2010, J. Chem. Theo. Comput., 6, 26, pMID: 26614316

\end{thebibliography}

\appendix

\section{QCT calculation data}
\begin{table*}[!ht]
\renewcommand{\arraystretch}{1.5}
\centering
\caption{Geometries (lengths in $\ang$ and angles in $^{\circ}$) of
  the stationary points for the \ce{HCl+(X^2$\Pi$) + H2(X^1$\Sigma$_g)
    -> H2Cl+(X^1A_1) + H($^2$S)} reaction. ``Ab intio'' refers to the UCCSD(T)-F12/ cc-pVQZ-F12 optimized values while ``PES'' refers to the fitted values.}
\begin{tabular}{|c|c|c|c|c|c|c|c|}
  \hline
Species&Level&$R_{\ce{HH^'}}$&$R_{\ce{HH^{''}}}$&$R_{\ce{H^{''}Cl}}$&$\Theta_{\ce{HH^{'}Cl}}$&$\Theta_{\ce{H^{''}ClH^{'}}}$&$\Phi_{\ce{HH^'ClH^{''}}}$\\

  \hline
 \ce{H2 + HCl+}&ab initio&0.74148&&1.31596&&&\\
\cline{2-8}
	&PES&0.74151&&1.31585&&&\\
 \hline
TS1	&ab initio&0.75063&2.06285&1.31670&82.48&41.22&81.34\\
\cline{2-8}
	&PES&0.75046&2.05911&1.31666&82.45&41.26&81.31\\
 \hline
M2	&ab initio&0.79532&2.39800&1.30530&79.03&87.34&89.36\\
\cline{2-8}
	&PES&0.79498&2.39768&1.30559&78.95&87.25&89.46\\
\hline
TS2	&ab initio&0.80189&2.48691&1.30501&99.13&88.31&84.33\\
\cline{2-8}
	&PES&0.80166&2.48873&1.30524&99.04&88.61&83.59\\
\hline
M3	&ab initio&1.38323&3.13620&1.30247&179.24&94.35&180.00\\
\cline{2-8}
	&PES&1.38748&3.14005&1.30261&179.22&94.38&180.00\\
 \hline
\ce{H + H2Cl+}&	ab initio&&&1.30554&&94.38&\\	
 \cline{2-8}
	&PES&&&1.30542&&94.38&\\
\hline	
\end{tabular}
\label{tab:theo_1}
\end{table*}

\begin{table*}[h]
\renewcommand{\arraystretch}{1.5}
\centering
\caption{Geometries (lengths in $\ang$ and angles in $^{\circ}$) of
  the stationary structures for the \ce{H^'H^{''}Cl+(X^1A_1) + H($^2$S)
    -> H^'($^2$S) + H^{''}HCl+(X^1A_1)} reaction. ``Ab intio'' refers to the the UCCSD(T)-F12/ cc-pVQZ-F12 optimized values.}
\begin{tabular}{|c|c|c|c|c|c|c|c|}
 \hline
Species&Level&$R_{\ce{H^'Cl}}$&$R_{\ce{H^{''}Cl}}$&$R_{\ce{HCl}}$&$\Theta_{\ce{H^{'}ClH^{''}}}$&$\Theta_{\ce{H^{''}ClH}}$&$\Phi_{\ce{H^'H^{''}ClH}}$\\
\hline
\ce{H + H2Cl+}&ab initio&1.30554&1.30554&&&94.38&\\	
\hline
TS3&ab initio&1.50695&1.29902&1.50705&89.02&89.02&180.00\\
\hline	
\end{tabular}
\label{tab:theo_2}
\end{table*}

\begin{table*}[h]
\renewcommand{\arraystretch}{1.5}
\centering
\caption{Relative energies (kcal/mol) and harmonic frequencies (cm$^{-1}$) of
  the stationary structures for the \ce{HCl+(X^2$\Pi$) + H2(X^1$\Sigma$_g) -> H2Cl+(X^1A_1) + H($^2$S)} reaction. ``Ab initio'' refers to the UCCSD(T)-F12/ cc-pVQZ-F12 optimized values while ``PES'' refers to the fitted values.}
\begin{tabular}{|c|c|c|c|c|c|c|c|c|}
 \hline
Species&Level&Energy&\multicolumn{6}{c|}{Frequency}\\
\cline{4-9}
&&&1&2&3&4&5&6\\
\hline
 \ce{H2 + HCl+}&ab initio&13.617&4403.8&2681.5&&&&\\	
 \cline{2-9}	
 &PES&13.599&4400.6&2679.7&&&&\\		
\hline	
 TS1	&ab initio&9.976&356.5i&4254.9&2662.0&534.1&351.1&211.4\\
 \cline{2-9}	
	&PES	&10.036&361.3i&4238.7&2655.5&538.6&348.5&195.8\\
\hline	
M2	&ab initio&3.272&3624.0&2757.0&675.8&588.1&385.5&234.6\\
 \cline{2-9}
	&PES&3.246&3614.4&2753.2&686.1&596.6&390.9&183.0\\
\hline
TS2	&ab initio&3.294&391.7i&3467.2&2744.9&662.5&497.9&199.6\\
 \cline{2-9}
	&PES&3.275&256.3i&3481.3&2755.7&702.0&624.3&369.1\\
\hline	
M3	&ab initio&	0.000 & 2779.6 & 1974.2& 1196.8 & 545.6 &401.8 & 325.6\\
 \cline{2-9}	
	&PES	&0.000 &2780.9&1987.9&1200.2 & 543.7 & 404.0 & 339.8\\
\hline	
\ce{H + H2Cl+}&ab initio & 3.852 & 2761.5 & 2752.0 & 1219.3 &&&\\
 \cline{2-9}					
	&PES         &3.846	& 2764.3	& 2751.9	& 1217.5	&&&\\			
\hline	
\end{tabular}
\label{tab:theo_3}
\end{table*} 

\begin{table*}[h]
\renewcommand{\arraystretch}{1.5}
\centering
\caption{Relative energies (kcal/mol) and harmonic frequencies (cm$^{-1}$) of
  the stationary structures for the \ce{H2Cl+(X^1A_1) + H($^2$S) ->
    H($^2$S) + H2Cl+(X^1A_1)} reaction. ``Ab intio'' refers to the UCCSD(T)-F12/ cc-pVQZ-F12 optimized values.}
\begin{tabular}{|c|c|c|c|c|c|c|c|c|}
 \hline
Species&Level&Energy&\multicolumn{6}{c|}{Frequency}\\
\cline{4-9}
&&&1&2&3&4&5&6\\
\hline
\ce{H + H2Cl+}&ab initio&0.0&2761.5&2752.0&1219.3&&&\\
\hline			
TS3&ab intio&14.547&1084.9i&2792.8&1634.0&1219.2&902.6&558.62\\
\hline	
\end{tabular}
\label{tab:theo_4}
\end{table*}

\begin{table*}[h]
\renewcommand{\arraystretch}{1.5}
\centering
\caption{Numbers of nonreactive and reactive trajectories, maximal impact
  parameter ($b_{max}$), and total reaction probability ($P$) for the
  \ce{HCl+(X^2$\Pi$) + H2(X^1$\Sigma$_g) -> H2Cl+(X^1A_1) + H($^2$S)}
  reaction in the QCT calculations. $N_{nonreative}$, $N_{H-abs}$,
  $N_{H-ex}$, and $N_{total}$ denote the numbers of the nonreactive,
  reactive H-abstraction, reactive H-exchange, and total trajectories, respectively.}
\begin{tabular}{|c|c|c|c|c|c|c|}
 \hline
  $T$~(K)&$b_{max}~(\ang)$&$N_{nonreative}$& $N_{H-abs}$& $N_{H-ex}$&  $N_{total}$&$P$\\
  \hline
300	&11.2&16665&4077&10&20752&$0.197 \pm 0.003$\\
  \hline
250	&11.4&16719&4405&7&21131&$0.209 \pm 0.003$\\
  \hline
200	&11.7&16544&4709&10&21263&$0.222 \pm 0.003$\\
  \hline
150	&12.0&16266&4932&18&21216&$0.233 \pm 0.003$\\
  \hline
100	&12.5&15791&5509&27&21327&$0.260 \pm 0.003$\\
  \hline
50	&13.1&16201&6571&42&22814&$0.290 \pm 0.003$\\
  \hline
20	&13.8&13960&7409&63&21432&$0.349\pm 0.003$\\
  \hline
\end{tabular}
\label{tab:theo_5}
\end{table*}

\end{document}